\documentclass{article}
\usepackage{layout}
\usepackage{graphicx} 
\usepackage{multicol}
\usepackage{multirow}
\usepackage{blindtext}
\usepackage{multicol}
\usepackage{geometry}
\usepackage{todonotes}
 \usepackage{afterpage}
\usepackage{lscape} 
\usepackage[utf8]{inputenc} 
\usepackage[T1]{fontenc}    
\usepackage{hyperref}       
\usepackage{url}            
\usepackage{booktabs}       
\usepackage{amsfonts}       
\usepackage{nicefrac}       
\usepackage{microtype}      
\usepackage{xcolor}         
\usepackage{amsmath} 
\usepackage{authblk}
\usepackage{float}
\usepackage{caption}  
\usepackage{bbm}

\usepackage{tabularx} 
\title{Diffusion Models in \textit{De Novo} Drug Design}

\author[1]{Amira Alakhdar \thanks{Email: aalakhda@andrew.cmu.edu}}
\author[2]{Barnabas Poczos \thanks{Email: bapoczos@cs.cmu.edu}}
\author[1, 3]{Newell Washburn \thanks{Corresponding author. Email: washburn@andrew.cmu.edu}}
\affil[1]{Department of Chemistry, Carnegie Mellon University, Pittsburgh, PA, 15213, USA}
\affil[2]{Department of Machine Learning, Carnegie Mellon University, Pittsburgh, PA, 15213, USA}
\affil[3]{Department of Biomedical Engineering, Carnegie Mellon University, Pittsburgh, PA, 15213, USA}

\date{\vspace{-5ex}}

\begin{document}

\maketitle
\begin{abstract}
\textbf{Diffusion models have emerged as powerful tools for molecular generation, particularly in the context of 3D molecular structures. Inspired by non-equilibrium statistical physics, these models can generate 3D molecular structures with specific properties or requirements crucial to drug discovery. Diffusion models were particularly successful at learning 3D molecular geometries' complex probability distributions and their corresponding chemical and physical properties through forward and reverse diffusion processes. This review focuses on the technical implementation of diffusion models tailored for 3D molecular generation. It compares the performance, evaluation methods, and implementation details of various diffusion models used for molecular generation tasks. We cover strategies for atom and bond representation, architectures of reverse diffusion denoising networks, and challenges associated with generating stable 3D molecular structures. This review also explores the applications of diffusion models in \textit{de novo} drug design and related areas of computational chemistry, such as structure-based drug design, including target-specific molecular generation, molecular docking, and molecular dynamics of protein-ligand complexes. We also cover conditional generation on physical properties, conformation generation, and fragment-based drug design. By summarizing the state-of-the-art diffusion models for 3D molecular generation, this review sheds light on their role in advancing drug discovery as well as their current limitations.}
\end{abstract}

\section{Introduction}
\begin{figure}[ht]
    \centering
    \includegraphics[width=1\linewidth]{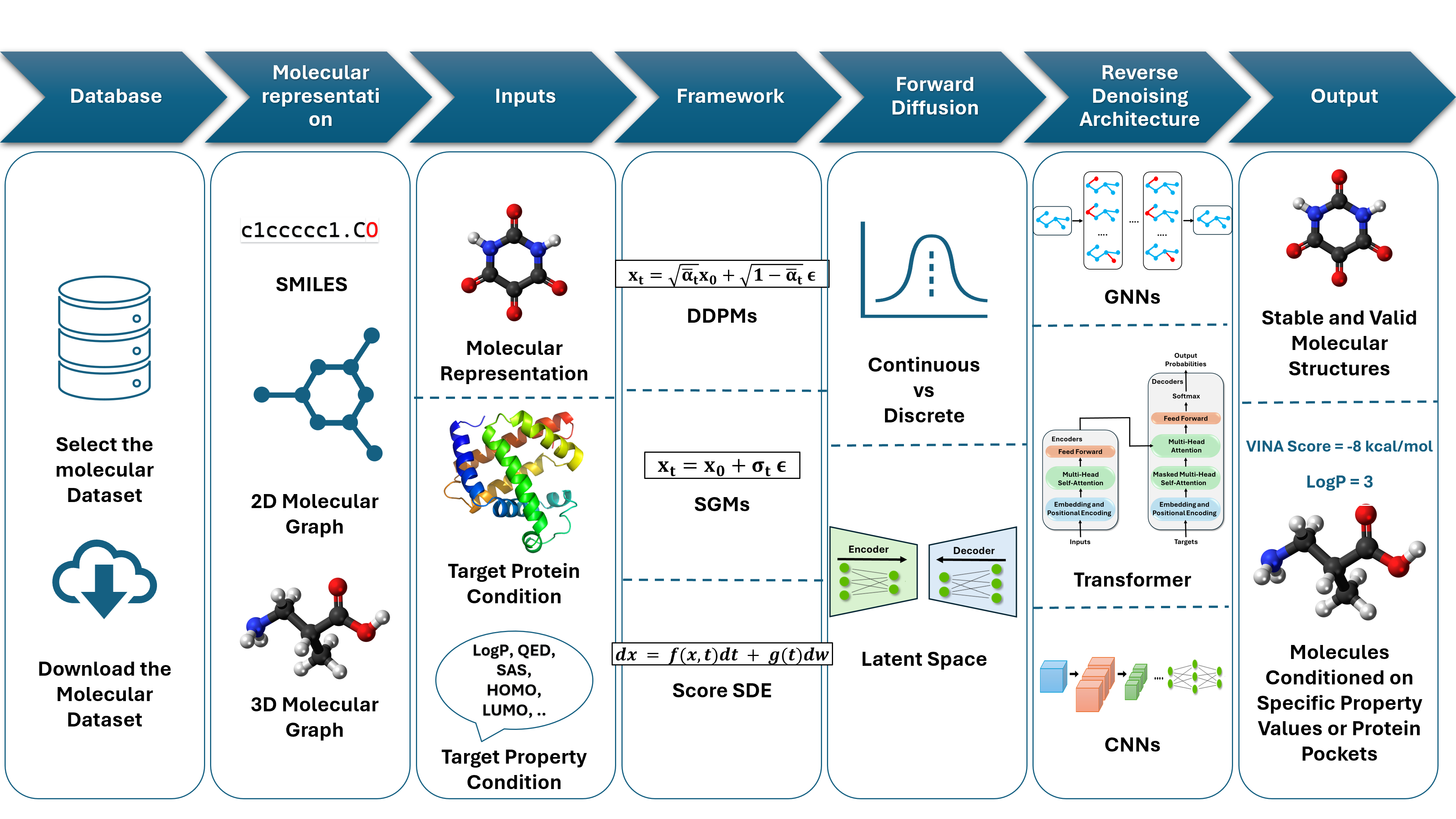}
    \caption{Overview of the process of generating molecules using diffusion models. First, the relevant dataset is acquired, the molecules are expressed in an appropriate molecular representation, and diffusion conditions are determined. Next, the diffusion framework (DDPM, SGM, Score SDE) is selected, and the forward and reverse diffusion strategies are designed. Denoising architectures may include transformers, GNNs, CNNs, and hybrid architectures. The output results are obtained, and the generated molecules are evaluated using multiple evaluation metrics according to the specific task in the drug discovery process [After\cite{Pang_Qiao_Zeng_Zou_Wei_2023}]. }
    \label{fig1}
\end{figure}

Generative models have been increasingly integrated with molecular science in modern drug discovery to help develop new therapeutics such as small molecules \cite{Pang_Qiao_Zeng_Zou_Wei_2023}, antibodies \cite{10.1093/bib/bbac267}, gene therapy and mRNA vaccines \cite{Kim_Mousavi_Yazdi_Zwierzyna_Cardinali_Fox_Peel_Coller_Aggarwal_Maruggi_2024b}. The generation of small molecules in the form of strings is well-established \cite{Pang_Qiao_Zeng_Zou_Wei_2023}. However, the generation of 3D molecular structures is still lagging due to the complexity of the shapes of molecules and the E(3) and SE(3) equivariance requirements for any model, which hold the molecules identical under rotations and translations \cite{Baillif_Cole_McCabe_Bender_2023}. The 3D geometry is the main determining factor of the electrochemical properties of the molecule that consequently determines its pharmacology, pharmacokinetics, metabolism, toxicity, and immune response by governing its interaction with the biological target pocket as well as various biological molecules such as (enzymes, receptors, antibodies, etc.). Hence, learning molecules in the 3D space can improve probing of the chemical space in structure-based drug design applications for possible protein and DNA ligands. Moreover, it can help advance the material discovery by conditioning on specific structural or quantum mechanics properties of the molecules. 
\\
Historically, computational molecular design or \textit{de novo} drug design (DNDD) was first approached by simpler methods such as growth and evolutionary algorithms \cite{doi:10.1517/17460441.2010.497534}. However, with the advancement of the deep generative models, they have taken over and were used to generate molecules in 1D,  2D, and more recently, in the 3D space. Several deep learning (DL) architectures were deployed for that end, including recurrent neural networks (RNNs), variational autoencoder (VAE), reinforcement learning (RL), generative adversarial network (GAN), convolutional neural network (CNNs), and Graph Neural Networks (GNNs). RNNs were popularly used with text-based representations such as SMILES and SELFIES, where they run in an “autoregressive” way to predict the next token of the sequence representing the molecule. RNNs were also used to generate molecular graphs in GraphNet \cite{li2018learning}, MolRNN \cite{Li_Zhang_Liu_2018}, and MRNN \cite{popova2019molecularrnn}, and those sequence-based models were usually combined with more complex architectures such as RL, VAE, and CNNs \cite{Elton_Boukouvalas_Fuge_Chung_2019}. Moreover, VAE-based methods such as  GraphVAE \cite{simonovsky2018graphvae}, CGVAE \cite{liu2019constrained}, NeVAE \cite{samanta2019nevae}, GAN-based models such as MOLGAN \cite{decao2022molgan}, GNN-based models such as GraphINVENT \cite{Mercado_Rastemo_Lindelöf_Klambauer_Engkvist_Chen_Bjerrum_2020}, and flow-based models such as GraphNVP \cite{madhawa2019graphnvp}, GraphAF \cite{shi2020graphaf}, MoFlow \cite{Zang_2020} and GraphDF \cite{luo2021graphdf} were used to generate the adjacency matrix of the 2D graph and in some cases, the 3D atomic coordinates. Several algorithms were developed specifically to generate the molecules in the 3D space, including autoregressive models or latent representation-based models that can be decoded from a VAE, an equivariant normalizing flow, or a diffusion process, and they are nicely summarized in a recently published review by Baillif at al \cite{Baillif_Cole_McCabe_Bender_2023}. Other reviews have focused on the implementation details for those models up to May 2022 \cite{Xie_Wang_Li_Lai_Pei_2022}; however, given that the advancement of diffusion models in generating 3D molecular structures has advanced quickly after that, most of the models with unprecedented results were not included in that review.\\
Inspired by non-equilibrium statistical physics, diffusion models were first introduced in 2015 by  Sohl-Dickstein et al to learn complex probability distributions through forward and reverse processes \cite{sohldickstein2015deep}. However, they became very popular in 2020 after achieving unprecedented results in image generation \cite{ho2020denoising}, and many studies followed to improve on those results \cite{kingma2023variational, nichol2021improved, bao2022analyticdpm} including the score-based method that was introduced by Song et al \cite{song2021scorebased}. Diffusion models also became popular in graph generation and they were used for several applications in computational biology and chemistry such as confirmation generation, molecular docking, protein generation and modeling, protein-ligand complex structure prediction, molecular generation, and molecular dynamics \cite{Guo_Liu_Wang_Chen_Wang_Xu_Cheng_2023, liu2023generative}. The first deployment of diffusion models for 3d molecular generation was the E(3) Equivariant Diffusion Model (EDM) \cite{hoogeboom2022equivariant} where they used an  E(n) equivariant graph neural network (EGNN) developed originally for discriminative tasks to denoise and learn molecular structures distributions \cite{satorras2022en}. Subsequently, they became widely adapted for 3D molecular generation tasks combined with GNNs or transformer-based models for encoding and learning molecule structures \cite{https://doi.org/10.13140/rg.2.2.26493.64480}. \\
In this review, we aim to summarize the technical implementation aspects of the diffusion models used for 3D molecular generation in general and for specific applications in drug discovery, such as \textit{de novo} drug design. There are several reviews in the literature about the use of deep generative models for \textit{de novo} drug design in general, and one of the very comprehensive examples is a recently published review by Xie et al. \cite{Pang_Qiao_Zeng_Zou_Wei_2023}. There are also several reports on the applications of diffusion models in bioinformatics and computational biology \cite{Guo_Liu_Wang_Chen_Wang_Xu_Cheng_2023, liu2023generative}. However, the literature lacks a review that dives into the recently developed diffusion models and compares their performance, evaluation methods, and implementation details. A related survey covers diffusion model applications in chemistry, including drug discovery until April 2023 \cite{https://doi.org/10.13140/rg.2.2.26493.64480}. They also covered antibody design, protein design, and material design applications and the emerging challenges in the field. In this review, we dive deeper into the representation and encoding of atoms and bonds and how those models aimed to learn both in synchronicity during the reverse diffusion to avoid generating unstable molecules with inconsistent atoms and bond structures. We cover the different strategies used for forward diffusion, the architectures employed for reverse diffusion, and the various applications of those models in the drug discovery process. An overview of the molecular generation process using diffusion models is shown in Figure.~\ref{fig1}.

\section{Diffusion Models}
\begin{figure}[ht]
    \centering
    \includegraphics[width=1\linewidth]{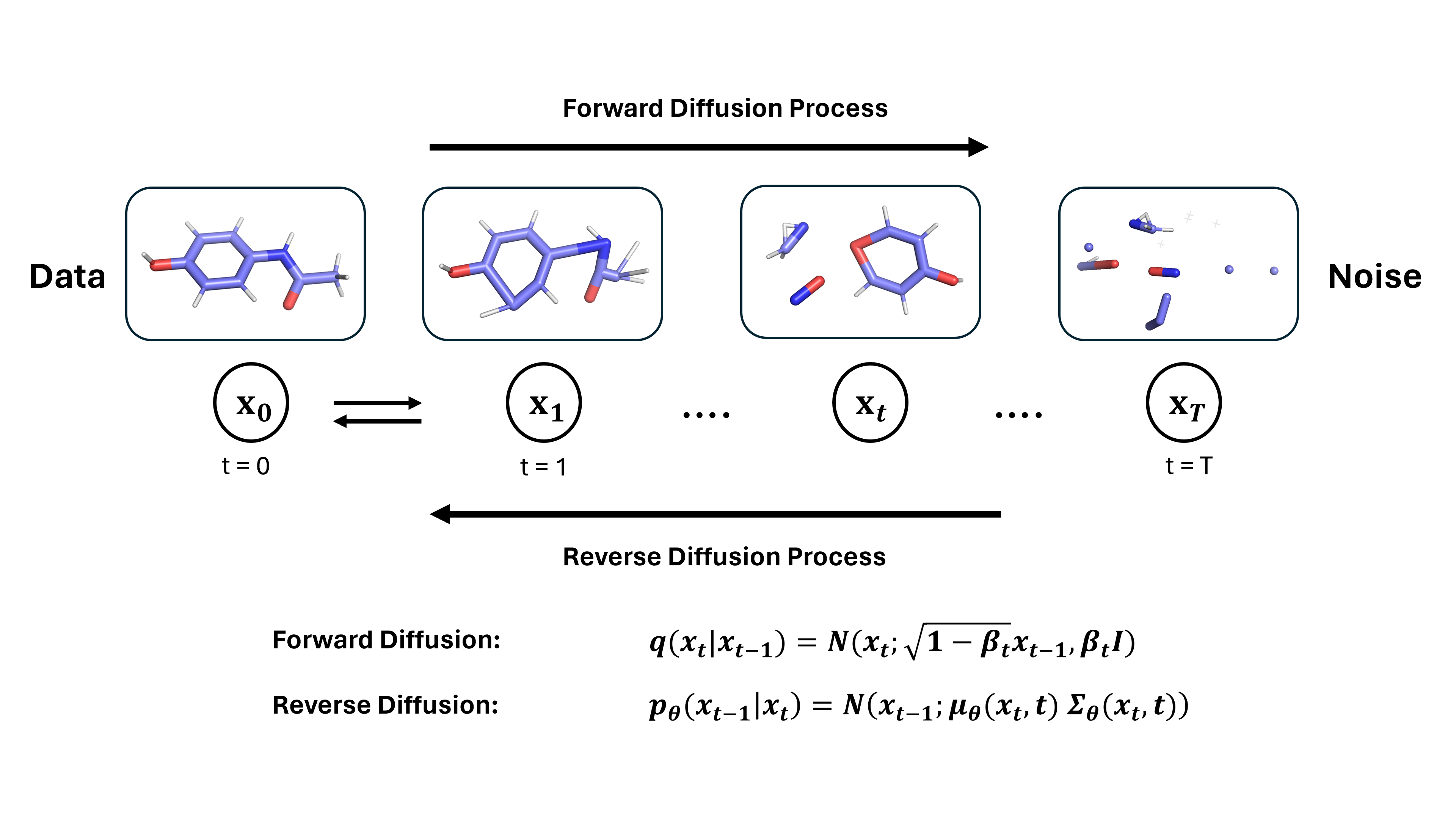}
    \caption{Overview of the diffusion process applied to 3D molecules. In the forward diffusion process, noise is added gradually to molecules by sampling from the distribution $\mathbf{q(x_t | x_{t-1}) = \mathcal{N}(x_t; \sqrt{1-\beta_t}  x_{t-1}, \beta_t \mathbb{I})}$ where $\mathbf{\beta_t \in (0, 1)}$ is a hyperparameter specified before model training, $\mathbf{\mathbb{I}}$ is the identity matrix and $\mathbf{t \in \{1, 2, \dots, T\}}$ is the time step. To generate molecules, starting from standard normal noise $\mathbf{x_T}$, samples are drawn from the distributions $\mathbf{p_{\theta}(x_{t-1}|x_t)}$ iteratively. Those distributions are learned by the pretrained denoising neural networks.}
    \label{fig2}
\end{figure}

Diffusion models are probabilistic generative models that add noise to distort the data and then reverse the process to generate samples. Current diffusion model research revolves around three main formulations: denoising diffusion probabilistic models (DDPMs), score-based generative models (SGMs), and models motivated by stochastic differential equations (score SDEs). Research on diffusion models focuses on improving several aspects of those three formulations, such as faster and more efficient sampling, accurate likelihood and density estimation, and handling data with special structures (e.g., permutation invariance, manifold structures, discrete data) \cite{yang2024diffusion}. The most popular formulation for molecular generations and DNDD applications is DDPMs, but other formulations are also used \cite{liu2023generative}. An illustration of the diffusion process applied to a 3D molecular graph is shown in Figure. \ref{fig2}. \\

\subsection{Denoising Diffusion Probabilistic Models (DDPMs)}
The denoising diffusion probabilistic model (DDPM) \cite{ho2020denoising, sohldickstein2015deep} incorporates two Markov chains: a forward chain that transforms data into standard Gaussian noise and a reverse chain that converts noise back to data by learning denoising transformations parameterized by deep neural networks. The generation of new data points from the same distribution can be achieved by sampling a random vector from the Gaussian distribution, followed by employing the reverse chain for ancestral sampling.
In formal terms, given a data distribution $\mathrm{x}_0 \sim q(\mathrm{x}_0)$,  the forward Markov process will use a Gaussian perturbation transition kernel $q(\mathrm{x}_t| \mathrm{x}_{t-1})$ to incrementally transform the data distribution $q(\mathrm{x}_0)$ into a tractable prior distribution $q(\mathrm{x}_T)$ where $T$ is the number of time steps as: 
\begin{equation}\label{eqn:1}
    q(\mathrm{x}_t | \mathrm{x}_{t-1}) = \mathcal{N}(\mathrm{x}_t; \sqrt{1-\beta_t} \mathrm{x}_{t-1}, \beta_t \mathbb{I}).
\end{equation}
Here $\beta_t \in (0, 1)$ is a hyperparameter specified before model training, $\mathbb{I}$ is the identity matrix and $t \in \{1, 2, \dots, T\}$ is the time step. With $\alpha_t := 1 - \beta_t$, and $\bar{\alpha_t} := \prod_{s = 0}^{t} \alpha_s$, it is easy to see that we can calculate $\mathrm{x}_t$ directly from $\mathrm{x}_0$ using the Gaussian kernel:
\begin{equation}\label{eqn:2}
    q(\mathrm{x}_t | \mathrm{x}_{0}) = \mathcal{N}(\mathrm{x}_t; \sqrt{\bar{\alpha_t}} \mathrm{x}_0, (1 - \bar{\alpha_t}) \mathbb{I})
\end{equation}
This process is followed by a learnable reverse denoising chain that reverses the time step and learns to retrieve the data gradually. The reverse chain takes the prior distribution $p(\mathrm{x}_T) \sim \mathcal{N}(0, \mathbb{I})$ and learnable transition kernel $p_{\theta}(\mathrm{x}_{t-1}| \mathrm{x}_t)$ such that:
\begin{equation}\label{eqn:3}
    q(\mathrm{x}_{t-1} | \mathrm{x}_{t}) = \mathcal{N}(\mathrm{x}_{t-1}; \mu_{\theta}(\mathrm{x}_t, t) \mathrm{x}_0, \Sigma_{\theta}(\mathrm{x}_t, t)) 
\end{equation}
where $\theta$ denotes model parameters, and the mean $\mu_{\theta}(\mathrm{x}_t, t)$ and variance $\Sigma_{\theta}(\mathrm{x}_t, t)$ are parameterized by deep neural networks. New samples can be generated by sampling from the Gaussian distribution $p(\mathrm{x}_T)$ and then iteratively sampling from the kernel in Eq.\ref{eqn:3} until we reach $t = 1$. Parameters of the neural network $\theta$ are learned by minimizing the Kullback-Leibler (KL) divergence between the distributions of the forward $q(\mathrm{x}_0, \mathrm{x}_1, \dots, \mathrm{x}_T )$ and the reverse $p_{\theta}(\mathrm{x}_0, \mathrm{x}_1, \dots, \mathrm{x}_T )$ Markov chains, which is equivalent to maximizing the variational lower bound (VLB) of the log-likelihood of the data $\mathrm{x}_0$ as shown in Eq.\ref{eqn:4}.
\begin{equation}\label{eqn:4}
\begin{aligned}
\mathbb{E}[-\log p_{\theta} (\mathrm{x}_0)] &\leq KL(q(\mathrm{x}_0, \mathrm{x}_1, \dots , \mathrm{x}_T) \| p_{\theta}(\mathrm{x}_0, \mathrm{x}_1, \dots , \mathrm{x}_T )) \\
&= - \mathbb{E}_{q(\mathrm{x}_0, \mathrm{x}_1, \dots , \mathrm{x}_T)}[\log p_{\theta}(\mathrm{x}_0, \mathrm{x}_1, \dots , \mathrm{x}_T )] + \text{const} \\
&=  - \mathbb{E}_{q(\mathrm{x}_0, \mathrm{x}_1, \dots , \mathrm{x}_T)}[- \log  p_{\theta} (\mathrm{x}_T) - \sum_{t=1}^{T} \log \frac{p_{\theta} (\mathrm{x}_{t-1} | \mathrm{x}_t)}{q(\mathrm{x}_t | \mathrm{x}_{t-1})}] + \text{const} \\
&= -\mathcal{L}_{VLB} (\mathrm{x}_0) 
\end{aligned}
\end{equation}
Ho et al. \cite{ho2020denoising} proposed an alternative $L_{VLB}$ loss by optimizing the neural network for predicting the Gaussian noise instead, as shown in Eq. \ref{eqn:5}.
\begin{equation}\label{eqn:5}
\mathbb{E}_{t \sim \mathcal{U} [1, T], \mathrm{x}_0 \sim q(\mathrm{x}_0), \epsilon \sim \mathcal{N}(0, 1)} [\lambda(t) \| \epsilon - \epsilon_{\theta} (\mathrm{x}_t, t) \|^{2} ]
\end{equation}
where $\lambda(t)$ denotes a positive weighting function, $x_t$ can be calculated as $\mathrm{x}_t = \sqrt{\bar{\alpha_t}} \mathrm{x}_0 +  \sqrt{1 - \bar{\alpha_t}} \epsilon$, $\epsilon_{\theta}$ is the neural network that will be trained to estimate the Gaussian noise, and $\mathcal{U} [1, T]$ is a uniform distribution over the set $\{1, 2, \dots, T \}$. \\
DDPMs were adopted in a wide variety of molecular diffusion models such as EDM \cite{hoogeboom2022equivariant}, MiDi \cite{vignac2023midi}, GCDM \cite{morehead2024geometrycomplete}, GeoDIff \cite{xu2022geodiff} and several others \cite{vignac2023digress, qiang2023coarsetofine, liu2023generative}.

\subsection{Score-Based Generative Models (SGMs)}
The concept of (Stein) score (also known as a score or score function) is a key quantity to score-based generative models. It is is defined as the gradient of the log probability density $\nabla_{\mathrm{x}} \log p(\mathrm{x})$ \cite{song2021scorebased, song2019sliced}. Score-based generative models (SGMs), also known as a noise-conditional score network (NCSN) \cite{song2020generative}, use a sequence of increasing Gaussian noise to perturb the data, then train a deep neural network model conditioned on noise levels to predict the score function. \\
To describe SGMs formally, we assume a data distribution $q(\mathrm{x}_0)$, and a sequence of increasing noise levels $0 < \sigma_1 <\sigma_2 < \dots < \sigma_t < \dots < \sigma_T$. The data will be perturbed from $\mathrm{x}_0$ to $\mathrm{x}_t$ by the Gaussian noise distribution $q(\mathrm{x}_t | \mathrm{x}_0) = \mathcal{N}(\mathrm{x}_t ; \mathrm{x}_0, \sigma_{t}^{2} \mathbb{I})$ to get the noisy data densities $q(\mathrm{x}_1), q(\mathrm{x}_2), \dots , q(\mathrm{x}_T)$ where $q(\mathrm{x}_t ) := \int q(\mathrm{x}_t | \mathrm{x}_0) q(\mathrm{x}_0) \mathrm{d}\mathrm{x}_0$, and the score function the score function $\nabla_{\mathrm{x}_t} \log q(\mathrm{x}_t)$ will be estimated using a noise-conditioned deep neural network $s_{\theta} (\mathrm{x}, t)$. Therefore, the objective loss function becomes:
\begin{equation}\label{eqn:6}
\begin{aligned}
&\mathbb{E}_{t \sim \mathcal{U} \|1, T\|, \mathrm{x}_0 \sim q(\mathrm{x}_0), \mathrm{x}_t \sim q(\mathrm{x}_t | \mathrm{x}_0)} \left[\lambda(t) \sigma_{t}^{2} \| \nabla_{\mathrm{x}_t} \log q(\mathrm{x}_t) - s_{\theta} (\mathrm{x}_t, t) \|^{2} \right]
\\
= &\mathbb{E}_{t \sim \mathcal{U} \|1, T\|, \mathrm{x}_0 \sim q(\mathrm{x}_0), x_t \sim q(\mathrm{x}_t | \mathrm{x}_0)} \left[\lambda(t)  \| - \frac{\mathrm{x}_t - \mathrm{x}_0}{\sigma_t} - \sigma_{t} s_{\theta} (\mathrm{x}_t, t) \|^{2} \right] + \text{const}
\\
= &\mathbb{E}_{t \sim \mathcal{U} \|1, T\|, \mathrm{x}_0 \sim q(x_0), \epsilon \sim \mathcal{N}(0, 1)} \left[\lambda(t)  \| \epsilon + \sigma_{t} s_{\theta} (\mathrm{x}_t, t) \|^{2} \right] + \text{const}
\end{aligned}
\end{equation}
Similar to Eq.\ref{eqn:5}, $\lambda(t)$ is a positive weighting function, and $x_t$ can be calculated as $\mathrm{x}_t = \mathrm{x}_0 +  \sigma_t \epsilon$. By comparing the loss functions in Eq.\ref{eqn:5} and Eq.\ref{eqn:6}, we can see that the training objectives of DDPMs and SGMs are equivalent. For sampling, starting from Gaussian noise, SGMs use a sequential chain of $ s_{\theta} (\mathrm{x}_T, T),  s_{\theta} (\mathrm{x}_{T-1}, T-1), \dots,  s_{\theta} (\mathrm{x}_0, 0)$ to produce new data instances. Annealed Langevin dynamics (ALD) is one of the most commonly used methods for sample generation in SGMs, but other methods, such as stochastic differential equations, ordinary differential equations, and their combinations with ALD have also been studied \cite{yang2024diffusion}. 

\subsection{Stochastic Differential Equations (Score SDEs)}
Score SDEs \cite{song2021scorebased} are an extension of DDPMs and SGMs to include infinite time steps or noise levels, with perturbation and denoising processes and they involve solving stochastic differential equations (SDEs) where SDEs are used for noise perturbation and sample generation, and the denoising is accomplished by estimating the score function of noisy data distributions. The data perturbation in Score SDEs-based diffusion is governed by the following SDE:
\begin{equation}\label{eqn:7}
     \mathrm{d} \mathrm{x} = \mathbf{f}( \mathrm{x}, t )\mathrm{d}t + g(t) \mathrm{d} \boldsymbol{\mathrm{w}}
\end{equation}
where $\mathbf{f}( \mathrm{x}, t )$ is the SDE diffusion function, $g(t)$ is the SDE drift function, and $\boldsymbol{\mathrm{w}}$ defines the standard Wiener process or Brownian motion. DDPMs and SGMs are both discretizations of this SDE in time, and the SDE of DDPM can be represented in \cite{song2021scorebased} as:
\begin{equation}\label{eqn:8}
     \mathrm{d} \mathrm{x} = - \frac{1}{2} \beta (t) \mathrm{x} \mathrm{d}t + \sqrt{\beta (t) }  \mathrm{d}  \boldsymbol{\mathrm{w}}
\end{equation}
Any diffusion process, taking the form of Eq. \ref{eqn:7} can be reversed by solving the following SDE \cite{anderson1982reverse}.:
\begin{equation}\label{eqn:9}
     \mathrm{d} \mathrm{x} = \left[ \mathbf{f}( \mathrm{x}, t) - g(t)^{2} \nabla_{\mathrm{x}} \log q_{t} (\mathrm{x})\right] \mathrm{d}t + g(t) \mathrm{d} \boldsymbol{\bar{\mathrm{w}}}
\end{equation}
Where $\boldsymbol{\bar{\mathrm{w}}}$ represents a standard Wiener process in the reverse-time diffusion process,  and $dt$ is an infinitesimal negative step in the opposite time direction. The solution of the reverse SDE has the same marginal densities as the forward SDE but in the opposite time direction. \\
After estimating the score function $\nabla_{\mathrm{x}} \log q_{t} (\mathrm{x})$ at each time step $t$, the reverse-time SDE (eq. \ref{eqn:9}) is solved and sampling can be achieved using numerical methods including numerical SDE/ODE solvers \cite{song2021scorebased, karras2022elucidating, song2022denoising, lu2022dpmsolver}, annealed Langevin dynamics \cite{song2020generative} and combination of MCMC with those methods (predictor-corrector methods) \cite{song2021scorebased}. Similar to SGMs, a time-dependent score neural network $s_{\theta} (\mathrm{x}, t)$ is trained to estimate the score function at each time step  using the following objective function in Eq. \ref{eqn:10} which is a generalization to the loss function in Eq. \ref{eqn:6} to continuous time,
\begin{equation}\label{eqn:10}
\mathbb{E}_{t \sim \mathcal{U} [0, T], \mathrm{x}_0 \sim q(\mathrm{x}_0), \mathrm{x}_t \sim q(\mathrm{x}_t | \mathrm{x}_0)} \left[\lambda(t) \sigma_{t}^{2} \| s_{\theta} (\mathrm{x}_t, t) - \nabla_{\mathrm{x}_t} \log q_{t}(\mathrm{x}_t | \mathrm{x}_0)   \|^{2} \right].
\end{equation}
Here $\mathcal{U} [0, T]$ denotes the uniform distribution over $[0, T]$. Similar to Eq.\ref{eqn:5} and Eq.\ref{eqn:6}, $\lambda(t)$ is a positive weighting function, and $\theta$ are the trainable parameters of the neural network. \\
Score-based models, including SGMs and Score SDEs, have been used in several diffusion models for molecule generation, such as DiffBridges \cite{wu2022diffusionbased}, GDSS \cite{jo2022scorebased}, and several others \cite{huang2023learning, lee2023exploring, liu2023generative}.

\section{Molecular Representations}
Over the years, there have been several ways to represent molecules such as coulomb matrices, molecular fingerprints, bag-of-bonds, International Chemical Identifier (InChI), Simplified molecular-input line-entry system (SMILEs), and graphs \cite{Pang_Qiao_Zeng_Zou_Wei_2023}, however, diffusion models have only been reportedly used with SMILEs, 2D and 3D graphs and achieved unprecedented results in molecular graph generation \cite{PENG2024122949, liu2023generative}. Hence, we will focus specifically on those two molecular data representations. Also, an overview of the datasets commonly used in molecular generation using the diffusion model is summarized in  Table. \ref{datasets_table}. 
\newline \\
\textbf{SMILEs} is a notation that translates a molecular structure into a one-dimensional string of symbols. Although sequence-based autoregressive models such as RNNs have achieved successful results with text-based representations, including SMILEs and SELF-referencIng Embedded Strings (SELFIES), diffusion models were investigated to apply SMILEs in more complex and tailored applications. For example, the DIFFUMOL model \cite{PENG2024122949} combined a diffusion model with Transformer architecture to tokenize SMILEs and generate molecules with specified scaffolds and properties.
\newline \\
\textbf{Molecular Graph Representation} has become a widely used representation in generative models where the molecular structure is represented as a graph $G = (V, E)$ with where nodes $v_i \in V$ represent the atoms and edges $(v_i, v_j) \in E$ represent the bonds or interatomic interactions between the atoms $v_i$ and $v_j$. In 2D graphs, bonds are represented as the edges, and atom types are represented as node features. In contrast, molecules represented in 3D graphs also have both the atom positions and atom types as node features, while the edges can be explicitly represented as edge features or implicitly encoded in the interatomic distances of the 3D atom coordinates while the edge features represent other information such as interatomic interactions in the fully connected graph as in the EDM model \cite{hoogeboom2022equivariant}. Both atom types and Bond types are encoded using one-hot embedding, where each channel represents an atom or bond type.

\begin{table}[H]
    \centering
    \renewcommand{\arraystretch}{1.5}
    \begin{tabularx}{\textwidth}{|p{3.1cm}|p{3cm}XX|}
    \toprule
    \textbf{Dataset} & \textbf{No. Molecules}  & \textbf{Data} & \textbf{Task} \\
    \toprule
    \textbf{QM9}  \cite{QM9} & 133,885 & Organic molecules with up to nine atoms and their DFT calculated quantum chemical properties. & Unconditioned and conditioned 3D molecular generation.\\
    \hline
    \textbf{GEOM-DRUG} \cite{GEOM} & over 450,000  &  37 million molecular conformations for over 450,000 molecules. & Unconditioned and conditioned 3D molecular generation.\\
    \hline
    \textbf{ZINC250k} \cite{ZINC} & 249,455 & A drug-like subset of the ZINC database with bioactivity data for a portion of the molecules. & Unconditioned and conditioned generation of drug-like molecules.\\
    \hline
    \textbf{MOSES} \cite{MOSES} & 1,936,962 & filtered from ZINC Clean Leads collection. & Unconditioned and conditioned generation of 2D molecules.\\
    \hline
    \textbf{CrossDocked} \cite{crossdock} &18,450 complexes, 22.6 million poses & A collection of ligand-protein complexes where ligands are docked against several protein targets similar to one another. &  SBDD tasks: target-aware 3D generation, molecular docking, etc.\\
    \hline
    \textbf{PDBbind} \cite{doi:10.1021/acs.accounts.6b00491} & 23,496 complexes & Biomolecular complexes in PDB with experimentally measured binding affinities, including 19,443 protein-ligand, 2,852 protein-protein, 1,052 protein-nucleic acid, and 149 nucleic acid-ligand complexes. & protein-protein docking and SBDD tasks: target-aware 3D generation, molecular docking, etc. \\
    \bottomrule
    \end{tabularx}
    \caption{Datasets commonly used in molecular generation using diffusion models}\label{datasets_table}
\end{table}

\section{Essential Requirements for Diffusion Models in Molecular Graph Generation}
\textbf{E(3) Invariance and SE(3) Equivariance:}
The model is considered E(3) equivariant when it is invariant to translations, rotations, and reflections of the 3D structure of the molecule. At the same time, SE(3) are only invariant to rotations and translations, which means that they are sensitive to chirality, which alters the 3D geometry of a molecule \cite{adams2021learning}. The main advantage of graph-based representation is they can easily meet the E(3) equivariance requirement by combining with geometric graph neural networks (GNNs) such as EGNN \cite{satorras2022en} and the rEGNN introduced in MiDi model \cite{vignac2023midi} for rotational invariance and centering the molecule by setting the center-of-mass to zero for translational invariance. 
\newline \\
\textbf{Permutation Invariant Graph Generation:}
Graphs are permutation invariant, meaning that they remain unchanged by permutation actions such as changing the order of the rows or columns in the adjacency matrix or the order of nodes corresponding to atoms and edges corresponding to bonds in the molecular graph representation $G = (V, E)$. Therefore, permutation invariance is one of the main requirements for any graph-based generation, including molecular generation. Diffusion models overcame the permutation invariant graph generation issue, surpassing autoregressive models that exhibit dependence on the sequence of generated nodes \cite{huang2022graphgdp}. 
\newline \\
\textbf{Accounting for Discreteness:}
\textbf{Accounting for Discreteness:}
Another critical issue with the graph representation is that the atom-type and bond-type features are discrete, making it challenging to use Gaussian noise diffusion to train a denoising neural network to learn the distribution of the molecular structures. Various forward and reverse diffusion methods applied to molecular graphs will be discussed in sections 5 and 6, respectively.
\newline \\
\textbf{Capturing Underlying Data Distributions:}
The generative model needs to learn the underlying distribution of the data within the chemical space and generate various data points accurately representing that distribution, including the 2D chemical graphs and their conformation in the 3D space. The molecular graph (2D structure) dictates the distribution of atom and bond types and the scaffolds and functional groups within compounds, and the 3D confirmation represents more sophisticated distributions such as bond angles, dihedral angles, and cases of stereo-isomerism.
\newline \\
\textbf{Generated Sample Fidelity:} 
Ensuring the validity of the generated molecules can have so many levels, including the connectivity of the entire molecular graph, chemical stability, and adherence to established rules of molecular structure such as atom valency, possible ionic charges based on atoms’ groups within the periodic table, stable ring structure and tolerable levels of angle and torsional strains that can change according to the underlying data distribution. For example, a set of drug-like molecules should have different underlying distributions from more reactive transition states that occur during a chemical synthesis. In section. 6, we will discuss how different architectures were employed to improve the fidelity of the generated molecules.

\section{Forward Diffusion Process (Discrete vs Continuous)}
In forward diffusion, the data is corrupted by injecting noise gradually until it reaches standard Gaussian noise. Given the categorical variable nature of atom and bond types, they are represented as discrete features in the molecular graph after one-hot encoding. Suppose we have a categorical array $h$ representing the categories  ${c_1, \dots, c_d}$. In that case, it can be one-hot encoded to the array $h^{\text{onehot}}$ using the one-hot function $h_{i,j}^{\text{onehot}} = \mathbbm{1}_{h_i=c_j}$ and then noise can be applied to the array $h^{\text{onehot}}$ innforward diffusion. Yet, it can be tricky to apply Gaussian noise to those features and then denoise the graph and maintain the same distribution of those categories in the generated data. \\
Nonetheless, several models applied Gaussian noise to discrete features, including EDM \cite{hoogeboom2022equivariant} that applies the continuous diffusion to the one-hot encoded vector using a predefined noise scaling schedule $q(z_{t}^{(h)} | h) = \mathcal{N}(z_{t}^{(h)}; \alpha_t h^{\text{onehot}}, \sigma_{t}^{2} \mathbb{I})$, where a probability distributions parameter $p$ is defined to be proportional to the normal distribution integrated from $1 - \frac{1}{2}$ to $1 + \frac{1}{2}$:
\begin{align*}
p(h|z_{0}^{(h)}) = \mathcal{C}(h|p), \quad p \propto \int_{1 - \frac{1}{2}}^{1 + \frac{1}{2}} \mathcal{N}(\boldsymbol{u}; z_{0}^{(h)}, \sigma_0) du 
\end{align*}
Here $\mathcal{C}(h|p)$ is a categorical distribution, and $p$ is normalized to sum to one. This ensures that one category will be active in each row of $z_{0}^{(h)}$ and in practice, the parameters of the normal noise distribution are tuned so that it is guaranteed that the sampled class from the reverse diffusion process matches the original active atom type category.\\
Moldiff \cite{peng2023moldiff} follows a similar forward diffusion process. However, they add one more none-type to the original element space and all-atom types or bond types will gradually be perturbed to this new type. They call it the absorbing type because atom or bond types are gradually absorbed to this specific type at the end of the forward diffusion process. Therefore, this perturbed probability distribution will be the starting point of the reverse process. Also, the forward diffusion was carried out in two stages, focusing on perturbing bond types in the first stage and atom types and coordinates in the second stage. \\
Other models, such as GFMDiff \cite{xu2024geometricfacilitated}, apply Gaussian noises with learnable parameters controlling the strength of noises. Similarly, JODO \cite{huang2023learning} projects the noise through learnable sinusoidal positional embeddings, and then it’s used as a conditional feature in the denoising process. Several studies favor continuous diffusion because it provides classifier-free guidance, i.e., an explicit classifier is not needed to guide the generative process, leading to more efficient training and sampling. Continuous diffusion can also provide uncertainty modeling and efficient sampling algorithms with faster designs based on advanced ODE solvers \cite{dieleman2022continuous, chen2023analog}. \\
On the other hand, other research suggested that discrete graph diffusion can generate higher-quality samples with distributions closer to the original data distribution indicated by the lower Maximum Mean Discrepancy (MMD) values between the two distributions \cite{haefeli2023diffusion}.MMD is a distance metric between two probability distributions discussed formally in the evaluation metrics section below. Discrete graph diffusion was also introduced in some molecular generation models, such as DiGress \cite{vignac2023digress}, to get better marginal distributions of node and edge types during diffusion. Simply, discrete forward diffusion is defined in DiGress \cite{vignac2023digress} as a categorical distribution applied to each node and edge type using a transition probability matrix. Given a one-hot encoded nodes (atoms) matrix $\mathcal{X} \in R^{n \times a}$ where $n$ is the number of nodes (atoms), and $a$ is the number of atoms categories, and a one-hot encoded edge tensor $\mathcal{E} \in R^{n \times n \times b}$ where $b$ is the number of edge types including the absence of edge as a particular edge type. In that case, the transition probabilities can be defined by the matrices $[\mathbf{Q}^{t}_{\mathcal{X}}]{i}j = q(x^t = j |x^{t-1} = i)$ and $[\mathbf{Q}^{t}_{\mathcal{E}}]_{ij} = q(e^t = j|e^{t-1} = i)$, and hence noise can be applied to the graph $G^t = (\mathcal{X}^t, \mathcal{E}^t)$ by sampling each node and edge type from a categorical distribution:
\begin{align*}
 q(G^t |G^{t-1} ) = (\mathcal{X}^{t-1} \mathbf{Q}^{t}_{\mathcal{X}}, \mathcal{E}^{t-1} \mathbf{Q}^{t}_{\mathcal{E}}) \quad \text{and} \quad  q(G^t |G) = (\mathcal{X} \bar{\mathbf{Q}}^{t}_{\mathcal{X}}, \mathcal{E} \bar{\mathbf{Q}}^{t}_{\mathcal{E}})   
\end{align*}
Here $\bar{\mathbf{Q}}^{t}_{\mathcal{X}} = \mathbf{Q}^{1}_{\mathcal{X}} \dots \mathbf{Q}^{t}_{\mathcal{X}}$ and $\bar{\mathbf{Q}}^{t}_{\mathcal{E}} = \mathbf{Q}^{1}_{\mathcal{E}} \dots \mathbf{Q}^{t}_{\mathcal{E}}$. The same discrete forward diffusion process was also used in a subsequent model called MiDi \cite{vignac2023midi}. However, it was combined with continuous diffusion for the 3D coordinates corrupted with Gaussian noise. Also, MiDi followed an adaptive noise schedule specifically tuned to make the model predict the bond types and atom coordinates before the atom types in the denoising process \cite{vignac2023midi}. 

\subsection{Latent Graph Diffusion}
\begin{figure}[ht]
    \centering
    \includegraphics[width=1\linewidth]{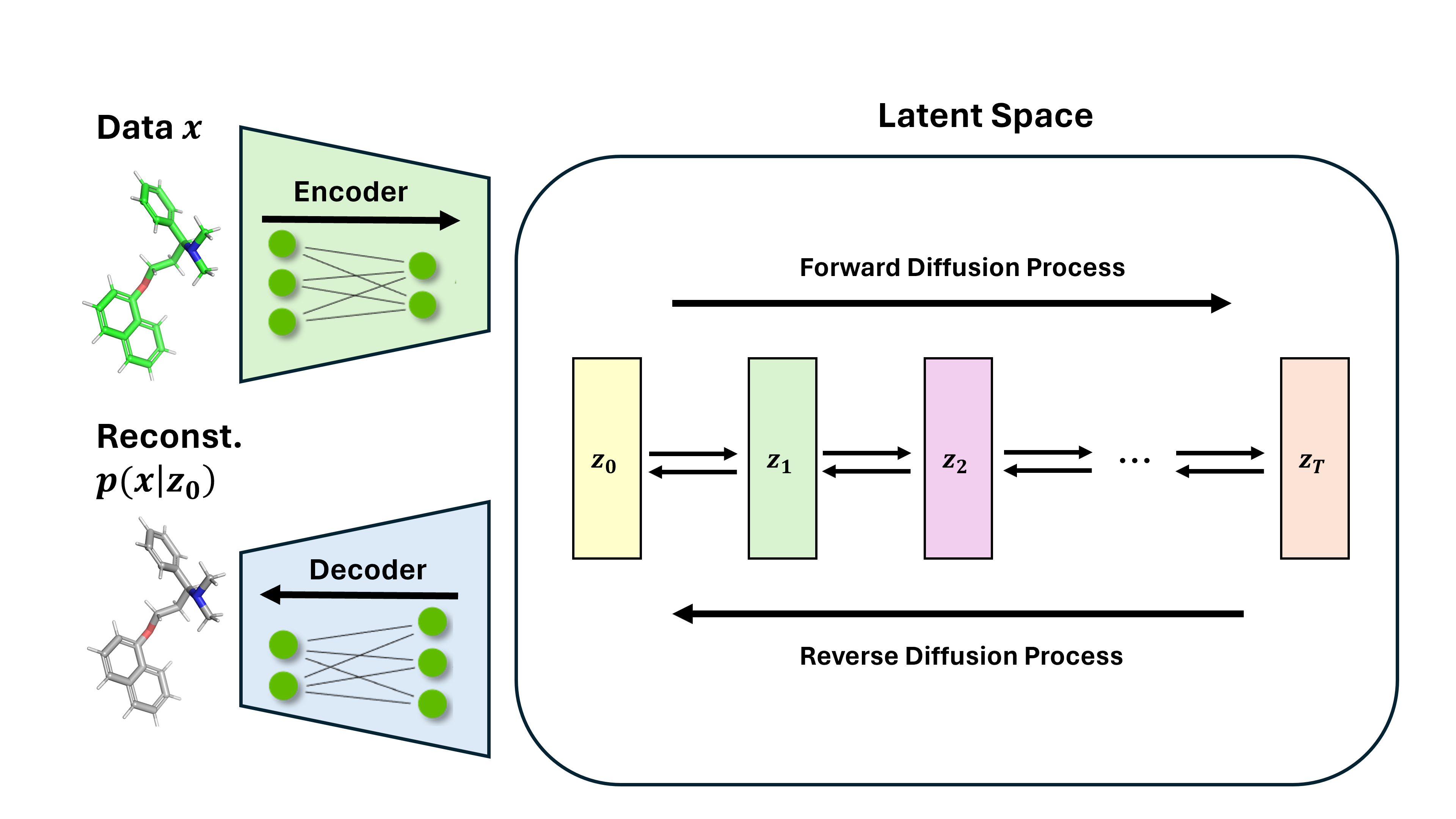}
    \caption{Overview of latent space diffusion process. First, the molecules are encoded to a continuous latent space, then stable diffusion is applied on the latent space. To generate molecules, they are first sampled from the latent space, then retrieved to the original discrete space using the decoder [After \cite{xu2023geometric}].}
    \label{fig3}
\end{figure}

Some models tend to embed the data in continuous space to apply stable diffusion, such as GEOLDM \cite{xu2023geometric}, which uses geometric autoencoders to convert molecular structures into 3D equivariant latent features and then apply stable diffusion in the latent space. Another model called 3M-Diffusion \cite{zhu20243mdiffusion} used a graph encoder to encode molecular graphs to a continuous space (latent space) aligned with their corresponding text descriptions, and then employs a decoder retrieve the molecular graph based on a given text descriptions. Hence, the model can simultaneously learn molecular structures and their text description and generate novel molecules based on a given textual description. An illustration of latent space diffusion is shown in Figure. \ref{fig3}.

\section{Reverse Diffusion Denoising Neural Networks Architectures}
\begin{figure}[ht]
    \centering
    \includegraphics[width=1\linewidth]{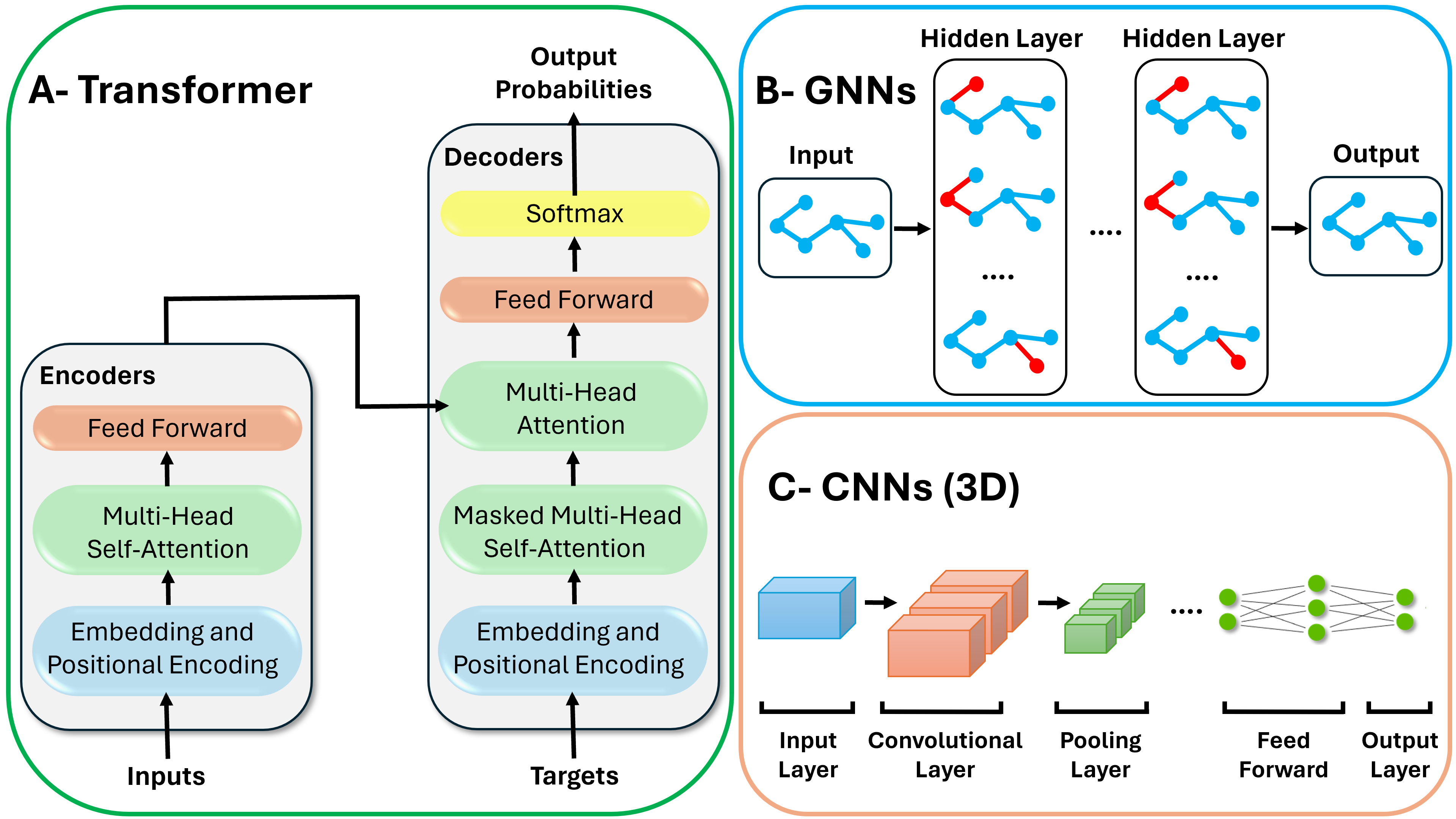}
    \caption{Simple illustrations of the neural network architectures commonly used in reverse diffusion: A- A-transformers, B-GNNs, and C-CNNs (in 3D).}
    \label{fig4}
\end{figure}

The reverse diffusion is responsible for learning the distribution of the data using a denoising neural network architecture that gradually removes the noise added in the forward diffusion.  Several architectures of denoising neural networks have been investigated for molecular generation in combination with diffusion models, and those architectures fall into one of the three categories that will be discussed in this section: transformers, GNNs, and CNNs (Figure. \ref{fig4} and Table. \ref{denoising_table}) or a combination of them. EDM also managed to generate molecules conditioned on prosperity $c$ by adding the condition as an input from which the EGNN could learn.

\subsection{Graph Neural Networks (GNNs)}
GNNs \cite{zhou2021graph} is a type of deep learning architecture specifically designed to work with data structured as graphs by passing messages between nodes, allowing each node to understand its neighbors and the broader network. Hence, they can operate on graph representations of molecules and have been widely adopted in predicting molecular properties and deep generative molecular models. Several variations of GNNs were also used in different models such as Graph Diffusion via the System of Stochastic differential equations (GDSS) \cite{jo2022scorebased}, E(n) Equivariant Graph Neural Networks (EGNNs) \cite{satorras2022en}, Geometry-Complete Perceptron Neural Networks (GCPNET) \cite{morehead2023geometrycomplete} and ShapeMol \cite{chen2023shapeconditioned}. GNNs have also been used to encode the discrete graph structures of molecules in the 3M-Diffusion model \cite{zhu20243mdiffusion} where they employ Graph Isomorphism Networks (GINs) \cite{hu2020strategies} to map molecules to a continuous latent space.

\subsubsection{EGNNs}
The EGNNs model developed originally for discriminative tasks, is the most popular GNNs architecture used in molecular generation using diffusion models \cite{satorras2022en}. It was first used in the EDM model to generate 3D molecules by message-passing through several Equivariant Convolutional Layers (EGCL) \cite{hoogeboom2022equivariant}. In the EDM model, the EGNNs used a fully connected graph $G$ with nodes $v_i \in V$, and Each node $v_i$ has the coordinates $x_i$ as its features, and hence the covalent bonds were not stated explicitly but inferred from the 3D coordinates. Implicit representation of covalent bonds caused several issues because the model cannot learn the valency of different atom types and the constraint relations between adjacent atoms that govern the bond formation and distributions. This issue was more clear in datasets of large drug-like molecules such as the GEOM-DRUGs \cite{GEOM} with up to 181 atoms and 44.4 atoms on average (24.9 heavy atoms) that generates more unstable molecules compared to the QM9 dataset \cite{QM9} with up to 29 atoms (9 heavy atoms) and an average of 18 atoms per molecule. EDM also managed to generate molecules conditioned on prosperity $c$ by adding the condition as an input from which the EGNNs can learn. \\
Several attempts have been made to improve various aspects of the EDM model relying on the same EGNN with modified models. For example, EEGSDE \cite{bao2023equivariant} used an SDE framework and added the guidance of energy function according to consistency between the molecule $z$ and the property $c$, which can help capture the dependency between the two variables and generate molecules with desired properties. DiffBridges \cite{wu2022diffusionbased} used a similar approach by designing physically informed diffusion bridges based on a \textit{Lyapunov function method}. An energy function was introduced to the diffusion bridge inspired by the AMBER force field \cite{ponder2003force} and molecular geometric statistics such as bond lengths, bond angles, torsional angles, etc, calculated from the data. The DiffBridges achieved better results on the QM9 dataset than EDM. However, the issue with inconsistent atom types and bonds leading to unstable molecules remained. \\
The MolDiff \cite{peng2023moldiff} model was developed to solve this issue of atom-bond inconsistency in EGNNs-based reverse diffusion, and hence, a two-stage diffusion model was implemented. In the first stage, bond types are perturbed with higher noise than atom types and positions that are only slightly perturbed. Then, the noised atom types and positions are used to recover bonds from the atom's information, and by that, the model learns the valency of molecules by relating bond types to atom types and coordinates, and the slight noise adds robustness to the model. In the following time step, atoms are adjusted based on the recovered atoms before bonds and are perturbed again until all bonds are labeled as the absorbing type. In the second stage, atom types and coordinates are perturbed until they reach the prior distribution of the none-type. Moldiff generated a higher percentage of valid molecules than EDM; however, it did not report on the stability of the generated 3D structures and atoms' valencies. \\
HierDiff \cite{qiang2023coarsetofine} was designed to generate higher-quality drug-like molecules, and it followed a completely different strategy to solve the atom-bond inconsistency issue in local environments using a hierarchical coarse-grained generation model where each node encodes a fragment. First, a vanilla EGNN obtains the embeddings for all links and nodes, and then the nodes are assembled to generate the 3d molecule using bottom-up or top-down EGNN that predicts the links between those fragments. The coarse-grained nodes are subsequently decoded into fine-grained fragments using another two EGNN modules: a message passing bottom-up EGNN and an iterative refined sampling bottom-up EGNN module. Similarly, GEOLDM used EGNN modules to implement the encoder, latent diffusion, and decoder \cite{xu2023geometric}. HierDiff and GEOLDM avoided the atom-bond inconsistency issue by embedding the structure into smaller domains using EGNNs.

\subsubsection{GCPNET of the GCDM}
GCPNET \cite{morehead2023geometrycomplete} is another SE(3)-equivariant graph neural network designed for discriminative tasks. It showed superior performance in 3D geometry-dependant predictive tasks on 3D molecular graphs such as protein-ligand binding affinity and chirality recognition on the rectus/sinister (RS) 3D molecular dataset \cite{adams2021learning}. It was repurposed for generative tasks in the GCDM model \cite{morehead2024geometrycomplete}. Like EGNNs, message passing is carried out through geometry-complete graph convolution layers called \textbf{GCPConv}; however, those layers are preceded by a Geometry-Complete Perceptron embedding layer $\textbf{GCP}_{e}$ to encode the input node and edge features into scalar and vector-valued values. The GCPConv layers are also followed by another GCP that outputs the final predictions of $\textbf{GCP}_{p}$. Although GCPNET has several advantages over EGNN, such as supporting geometry-complete and chirality-aware message-passing, it failed to improve the 3d stability of larger molecules such as the GEOM-DRUGs dataset \cite{GEOM} in the GCDM model \cite{morehead2024geometrycomplete}.

\subsubsection{ShapeMol}
ShapeMol generates 3D molecules conditioned on molecular shape using two GNNs: an invariant graph neural network, denoted as INV-GNN, to predict atom features and an equivariant graph neural network, denoted as EQ-GNN to predict atomic coordinates \cite{chen2023shapeconditioned}.

\subsubsection{MACE of the SiMGen Model}
The MACE model \cite{batatia2023mace} is an equivariant message-passing neural networks (MPNNs) architecture designed to create computationally efficient and accurate ML-based force fields. Hence, it was used to extract features from the SPICE \cite{Eastman_Behara_Dotson_Galvelis_Herr_Horton_Mao_Chodera_Pritchard_Wang_et} dataset with 1 million molecules of sizes ranging from 3 to 100 atoms for the zero-shot molecular generation model, SiMGen \cite{elijošius2024zero}. The QM force field features were combined with time-varying local similarity Kernels to define a score-based diffusion model that can generate molecules of any arbitrary size and execute conditional generation without altering the model.

\subsubsection{LigandDiff}
LigandDiff \cite{doi:10.1021/acs.jctc.4c00232} is a conditional diffusion model aiming to generate 3D transition metal complexes employing message-passing neural networks (MPNNs) for denoising combined with Geometric vector perceptrons (GVPs) for embedding molecules into molecular representation. LigandDiff conditionally generates molecules under a fixed context; for example, the number of heavy atoms and the transition metal used can be used as a condition.

\subsection{Convolutional Neural Networks (CNN)}
CNNs are deep neural networks especially proficient at processing data organized into a grid-like structure, like images. CNNs are frequently utilized for computer vision applications, including object detection, image classification, and image recognition \cite{oshea2015introduction}. \\

\subsubsection{SchNet of the MDM Model}
SchNet \cite{schütt2017schnet} is a continuous-filter CNN that learns a representation of atoms in a molecule analogous to pixels in an image where atoms are embedded based on the atom type with three interaction blocks modeling the interatomic interactions. The model is equivariant and rotational invariant because it uses interatomic distances to model molecules in the filter network. SchNet \cite{schütt2017schnet} was adapted in the MDM model \cite{huang2022mdm} to generate 3D molecular structures. It was used to extract node embeddings along with Dual Equivariant Score Neural Networks: one operates on the molecular graph to tackle covalent bonding interactions using local edges, and one operates on the fully connected molecular graph to tackle interatomic van der Waals forces using global edges. MDM \cite{huang2022mdm} managed to improve the 3d stability of the GEOM-DRUGs \cite{GEOM} dataset compared to the EDM model. \\

\subsubsection{3D U-Net of the VoxMol model}
The VoxMol \cite{pinheiro20243d}, a score-based model, employed a CNN architecture designed for volumetric segmentation tasks called 3D U-Net for the denoising process. The 3D U-Net is an expansion of the wildly successful conventional U-Net architecture for 2D image segmentation \cite{ronneberger2015unet}. The atomic densities of atoms are depicted as continuous Gaussian-like values within a 3D grid space where each atom type (element) is represented by a different grid channel. Molecules are then generated by discretizing the 3D space around the atoms into voxel grids. Voxelized molecules can later be recovered from the generated voxel grids using a simple peak detection algorithm. Given the voxelized representation of molecules, the CNNs-based VoxMol model was able to achieve remarkable results with the GEOM-DRUGs dataset \cite{GEOM} in terms of 3D molecular stability.

\subsection{Transformers}
Transformers are a type of neural network design that has gained popularity in natural language processing (NLP). Typical transformer architecture consists of multi-head attention (MHA), layer normalization (LN), and feed-forward networks (FFN). Their ability to capture long-range dependencies between word sequences using self-attention mechanisms allows them to analyze the relationships between different parts of a text input simultaneously  \cite{vaswani2017attention}. Similarly, they have shown a significant ability to capture long-range dependencies between atoms and bonds and, hence, learn the 3D molecular structures efficiently. Message-passing transformers can be tailored for graph-structured data like molecules and have been employed in several denoising architectures such as JODO \cite{huang2023learning} and MiDi \cite{vignac2023midi} that were highly successful in the 3D graph generation compared to the early EGNN-based models such as EDM \cite{hoogeboom2022equivariant}. Moreover, in the DIFFUMOL \cite{PENG2024122949} model, the transformer architecture was used to model the semantic relationship between the user-guided instructions and the generated molecules.

\subsubsection{Diffusion Graph Transformer (DGT) of the JODO Model}
The JODO model \cite{huang2023learning} employs a DGT for parameterizing the data prediction architecture. DGT adopts the typical transformer architecture; however, it interacts intricately with node (atoms) and edge (bonds) representations to learn the dependencies between them. DGT employs an adaptive layer normalization (AdaLN) \cite{Guo_Wang_Yu_McKenna_Law_2022} to project edge and atom features to continuous embeddings. Moreover, the model uses a self-conditioning mechanism \cite{chen2023analog} where predictions from the previous sampling step are utilized as an extra condition where a 2D adjacency matrix and a 3D adjacency matrix calculated from distance cut-offs are derived from the previous sampling step to serve as a condition for the ongoing sampling step, therefore enhancing the model’s capture the precise graph discreteness along with its connectivity and spatial arrangement dependencies on atom types. Besides self-conditioning, they used different methods to augment the data, such as the m-step random walk matrix from the discrete adjacency matrix and discrete atomic features. JODO achieved highly successful results with conditional and joint unconditional 2D and 3D molecular graph design \cite{huang2023learning}.

\subsubsection{Dual Track Transformer Network (DTN) of the GFMDiff Model}
DTN, the E(n) equivariant denoising kernel of the GFMDiff \cite{xu2024geometricfacilitated} model, is designed to specifically learn the 3D geometry of the molecules by taking embedding inputs derived from atom features, pairwise distance features, and triple-wise angle features, and predicts atom types, valencies, and coordinates as an output. The spatial information was used to learn multi-body interactions among atoms. Moreover, the GFMDiff model introduced a new loss function, the Geometric-Facilitated Loss (GFLoss), that takes the valencies predicted by the DTN and aims to minimize the difference between those valencies and valencies calculated from molecular geometries. This architecture of the DTN, combined with the GFLoss, demonstrated a notable efficiency in learning atomic valencies and achieved favorable molecular and atomic stabilities of the generated molecules.

\subsubsection{Molecule Unified Transformer (MUformer) of the MUDiff Model}
Similar to the DGT of the JODO model \cite{huang2023learning}, MUformer \cite{hua2024mudiff} is an E(n) equivariant transformer architecture that jointly learns the 2D and the 3D graph structures of the molecule. It relies on encoding atomic, positional, and structural information using six encoding functions, three of which are message-passing based. Those six channels generate atom, bond, and graph encodings besides 2D and 3D neighborhood encoding that combine into two channels: one that represents 2D molecular structures and predicts atoms and edge features and one that represents the 3D geometric structure and predicts the atom features and the 3D geometry. Finally, an output network merges the atom features from the two channels and outputs the final molecule. The MUformer successfully generated valid and stable 3D geometries and property prediction tasks on the QM9 dataset \cite{QM9}.

\subsubsection{DiGress}
DiGress employed a discrete denoising architecture to retrieve the 2D molecular graph as categorical atom and bond features. The denoising architecture is based on a graph transformer network from \cite{dwivedi2021generalization} that uses FiLM layers \cite{perez2017film} to combine edge features and global features. Similar to MUformer \cite{hua2024mudiff}, theoretic structural features such as cycles and spectral features, as well as molecular features such as the current valency of each atom and the current molecular weight of the whole molecule, are used to enhance the representations of molecules to improve the denoising process.

\subsubsection{MiDi}
MiDi also employs a graph-denoising transformer architecture that jointly predicts molecules' 2D and 3D graphs. MiDi allows molecule atoms to hold formal charges and become ions; a carbon atom, for example, can hold charges of -1, 0, and 1. The denoising architecture learns to predict the molecular structure gradually, where atom coordinates, and bond types are predicted first, followed by the atom types and formal charges. MiDi's denoising architecture integrates relaxedEGNN (rEGNN) layers into its update block. MiDi uses edge, node, pairwise, and global features pooled into node representations using the Principal Neighbourhood Aggregation (PNA) layers \cite{corso2020principal}. 

\subsection{Hybrid Architectures}
Combining elements from various architectures, such as transformers and GNNs, can leverage the strengths of each approach to enhance denoising performance. A good example of that is the Conditional Diffusion model based on discrete Graph Structures (CDGS) \cite{huang2023conditional}, where a hybrid message passing block (HMPB) was employed for the denoising architecture. HMPB includes two variations of message-passing layers: a GNN layer called GINE \cite{hu2020strategies} for discrete data types to aggregate local neighbor node-edge features and a fully connected attention-based transformer \cite{vaswani2017attention} for global graph features learning. \\
Similar to ShapeMol \cite{chen2023shapeconditioned}, Diff-Shape \cite{Lin_Xu_Chen_2024} proposed a pre-trained GNN called Graph ControllNet (GrCN) to condition the diffusion process on 3D molecular shapes. Graph ControllNet is inspired by ControlNet \cite{zhang2023adding}, which was pre-trained with billions of images to guide text-to-image diffusion models. Similarly, a pre-trained Graph ControllNet (GrCN) was combined with a pre-trained unconditioned model of MiDi \cite{vignac2023midi} to generate molecules constrained by 3d shape. While MiDi \cite{vignac2023midi} contains a transformer architecture, and GrCN is GNN-based, the model can be considered a hybrid architecture. Diff-Shape \cite{Lin_Xu_Chen_2024} can also be employed for structure-based drug design by conditioning on the 3D shape of ligands pre-docked to the target protein with high docking scores.

\afterpage{\clearpage}

\begin{table}[ht]
  \centering
  \renewcommand{\arraystretch}{1.5}
  \begin{tabularx}{\textwidth}{|p{2.0cm}||p{2.0cm}|p{2.0cm}Xp{1.8cm}p{4.2cm}|}
    \toprule
    Denoising Architecture     &  Model   & Denoising Model & Condition & Framework & Datasets \\
    \midrule
    \hline
    \multirow{11}{4em}{GNNs} &  EDM \cite{hoogeboom2022equivariant} & EGNNs   \cite{satorras2022en} & Conditioned, unconditioned & DDPM & QM9 \cite{QM9}, GEOM-Drugs \cite{GEOM}   \\ 
    & DiffBridges \cite{wu2022diffusionbased} &  EGNNs \cite{satorras2022en} & Unconditioned & SMLD$^a$  & QM9 \cite{QM9}, GEOM-Drugs \cite{GEOM} \\
    & EEGSDE \cite{bao2023equivariant}  & EGNNs \cite{satorras2022en} & Conditioned, unconditioned & Score SDE & QM9 \cite{QM9}, GEOM-Drugs \cite{GEOM}  \\
    & GEOLDM \cite{xu2023geometric} & EGNNs \cite{satorras2022en} & Conditioned, unconditioned & DDPM & QM9 \cite{QM9}, GEOM-Drugs \cite{GEOM} \\
    & Moldiff \cite{peng2023moldiff} & EGNNs \cite{satorras2022en}& Unconditioned & DDPM & QM9 \cite{QM9}, GEOM-Drugs \cite{GEOM} \\
    & HierDiff \cite{qiang2023coarsetofine} & EGNNs \cite{satorras2022en}& Conditioned & DDPM & GEOM-Drugs \cite{GEOM}, CrossDocked2020 \cite{crossdock}\\
    & GCDM \cite{morehead2024geometrycomplete}& GCPNET \cite{morehead2023geometrycomplete}&  Conditioned, unconditioned & DDPM & QM9 \cite{QM9}, GEOM-Drugs \cite{GEOM}, RS \cite{adams2021learning} \\
    & ShapeMol \cite{chen2023shapeconditioned} &   INV-GNNs, EQ-GNNs & Conditioned & DDPM & MOSES \cite{MOSES}\\
    & GDSS \cite{jo2022scorebased} & & Unconditioned & Score SDE & QM9 \cite{QM9}, ZINC250k \cite{ZINC} \\
    & LigandDiff \cite{doi:10.1021/acs.jctc.4c00232} & MMPNs, GVPs & Conditioned & DDPM & Subset of Cambridge Structural Database (CSD) \cite{Arunachalam_Gugler_Taylor_Duan_Nandy_Janet_Meyer_Oldenstaedt_Chu_Kulik_2022, Groom_Bruno_Lightfoot_Ward_2016} \\
    \hline
    \multirow{2}{4em}{CNNs} &  MDM \cite{huang2022mdm} & Schnet \cite{schütt2017schnet} &  Conditioned, unconditioned & SMLD$^a$  & QM9 \cite{QM9}, GEOM-Drugs \cite{GEOM} \\ 
    & VoxMol \cite{pinheiro20243d} & 3D U-Net \cite{ronneberger2015unet} & Unconditioned & SGM & QM9\cite{QM9}, GEOM-Drugs \cite{GEOM} \\
    \hline
    \multirow{6}{4em}{Transformer} &  DiGress \cite{vignac2023digress} & & Conditioned, unconditioned & DDPM  & QM9\cite{QM9}, GuacaMol \cite{Brown_2019}, MOSES \cite{MOSES} \\ 
    & MiDi \cite{vignac2023midi} & & Unconditioned & DDPM & QM9\cite{QM9}, GEOM-Drugs \cite{GEOM}  \\
    & DIFFUMOL \cite{PENG2024122949} & & Conditioned & DDPM & GuacaMol\cite{Brown_2019}, MOSES \cite{MOSES}, ZINC250K \cite{ZINC} \\
    & JODO \cite{huang2023learning} & DGT &  Conditioned, unconditioned  & Score SDE & QM9\cite{QM9}, GEOM-Drugs \cite{GEOM}  \\
    & MUDiff \cite{hua2024mudiff} & MUformer &  Conditioned, unconditioned & DDPM & QM9\cite{QM9} \\
    &  GFMDiff \cite{xu2024geometricfacilitated} & DTN & Conditioned, unconditioned & DDPM & QM9 \cite{QM9}, GEOM-Drugs \cite{GEOM} \\
    \hline
    \multirow{5}{4em}{Other} &  CDGS \cite{huang2023conditional}  & HMPB (hybrid) & Unconditioned & Score SDE & QM9 \cite{QM9}, ZINC250K \cite{ZINC} \\ 
    &  Diff-Shape \cite{Lin_Xu_Chen_2024} & hybrid & Conditioned & DDPM & GEOM-Drugs \cite{GEOM}, PDBBind dataset \cite{doi:10.1021/acs.accounts.6b00491} \\ 
    & SiMGen \cite{elijošius2024zero} & & Conditioned & Score SDE & QM9 \cite{QM9}, SPICE \cite{Eastman_Behara_Dotson_Galvelis_Herr_Horton_Mao_Chodera_Pritchard_Wang_et}\\
    & 3M-Diffusion \cite{zhu20243mdiffusion} & & Conditioned & DDPM & ChEBI-20 \cite{Hastings_Owen_Dekker_Ennis_Kale_Muthukrishnan_Turner_Swainston_Mendes_Steinbeck_2015}, PubChem \cite{liu-etal-2023-molca}, PCDes \cite{Zeng_Yao_Liu_Sun_2022}, MoMu \cite{su2022molecular} \\
    & GSDM \cite{luo2022fast} & & Unconditioned & Score SDE & QM9 \cite{QM9}, ZINC250k \cite{ZINC} \\
  \bottomrule
  \hline
  \end{tabularx}
  \caption{Summery of the denoising architectures used in diffusion models for molecular generation, and the datasets used for training those models.}
  \label{denoising_table}
  \footnotesize{$^a$ SMLD: score matching with Langevin dynamics, a sub category of SGMs where Langevin dynamics are used}
\end{table}

\section{ Molecular Generation for Structure-Based Drug Design (SBDD)}
\begin{figure}[ht]
    \centering
    \includegraphics[width=1\linewidth]{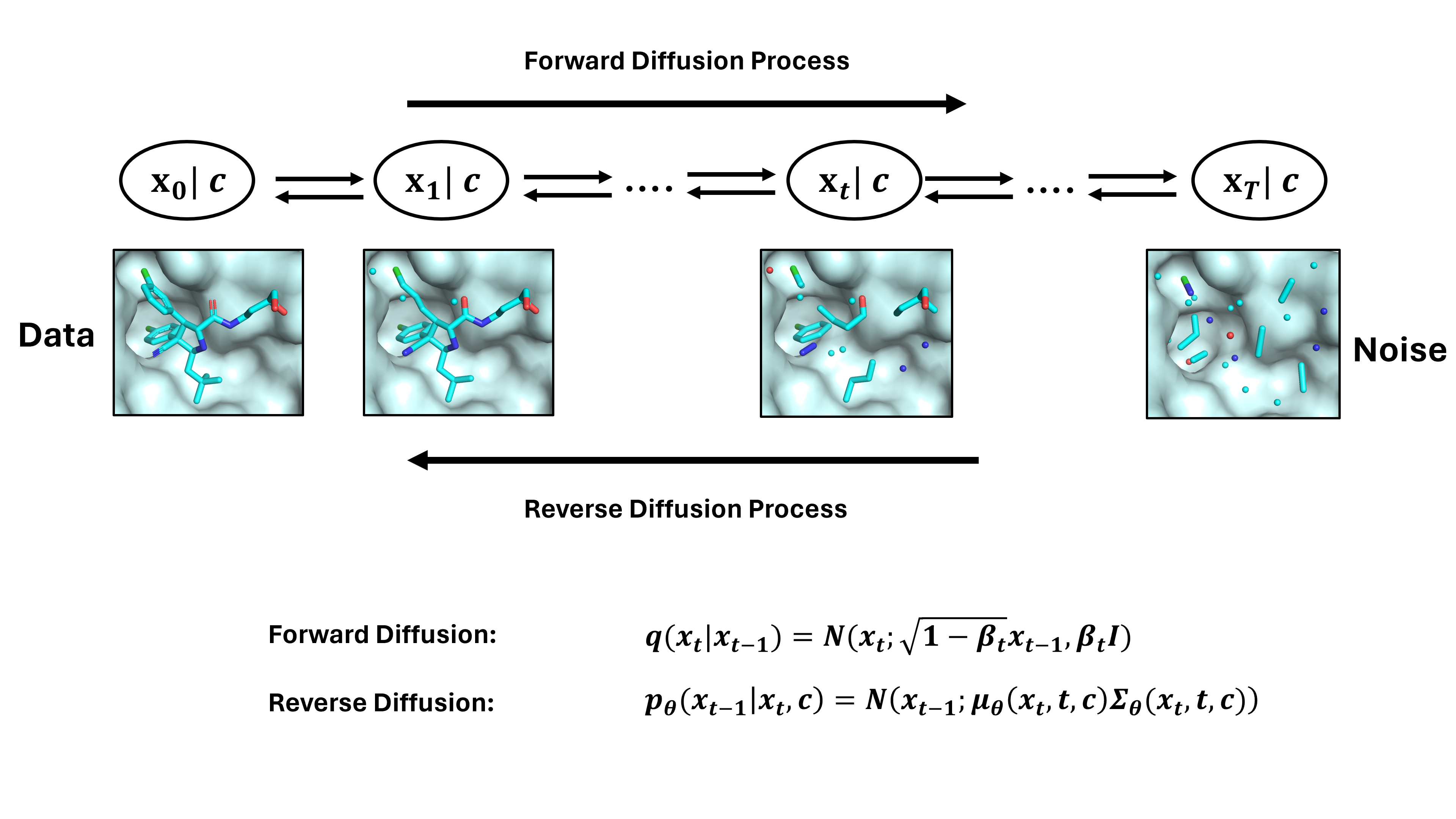}
    \caption{Generation of 3D molecules conditioned on protein pocket using diffusion models.}
    \label{fig5}
\end{figure}

Molecular generation using diffusion models can be guided using the protein target binding pocket as a condition for the generation. In the case of 3D generation, the model can also generate the pose of the molecule within the pocket (Figure. \ref{fig5} and Table. \ref{applications_table}). DiffSBDD \cite{schneuing2023structurebased} and DiffBP \cite{lin2022diffbp} were the first two models to introduce SBDD-based diffusion models, and they both used an equivariant EGNN for modeling the molecules; however, the DiffSBDD \cite{schneuing2023structurebased} study was later expanded to explore more cases. Currently, it includes two different approaches. In the first approach, a fixed pocket representation is used as a condition for the 3D molecular generation, while in the second approach, the joint distribution of ligand-pocket complexes was unconditionally approximated. They also studied the impainting strategies where the context can be injected into the sampling process at the probabilistic transition steps. Impainting allows for masking, replacing, or fixing arbitrary parts of the ligand-pocket system. Hence, it can be combined with the second strategy for full 3D molecular generation or fixing parts of the molecule and performing partial molecular generation or fragment generation. This can allow for optimizing candidate drug molecules (leads) by exploring the local chemical space while maintaining the rest of the 3D molecular structure, as well as for applying techniques widely used in drug design, such as fragment generation, scaffold hopping, fragment growing, and linker design. \\
TargetDiff \cite{guan20233d} is another target-aware molecular generative model that employs graph attention layers for parameterization. It also studied the case of sub-structure or fragment generation and binding affinity ranking and prediction. Both models were able to achieve better Vina Scores \cite{10.1002/jcc.21334} on the CrossDocked2020 dataset \cite{crossdock} than state-of-the-art non-diffusion-based models such as Pocket2Mol \cite{peng2022pocket2mol} and GraphBP \cite{liu2022generating}. However, their performance needs more improvement regarding other drug design metrics, such as SAS and QED scores. \\
Moreover, two models relying on the idea of decomposing ligands were developed: DECOMPOPT \cite{zhou2024decompopt} and DECOMPDIFF \cite{guan2024decompdiff} where ligands are decomposed into two parts arms, and scaffold with arms are responsible for forming interactions with the pocket amino acid residues and the scaffold connects all arms to form a complete molecule. DECOMPDIFF \cite{guan2024decompdiff} uses structural priors over arms and scaffold, including priors estimated from the reference molecule and pocket prior derived from the subpockets within the target binding site extracted by AlphaSpace2 \cite{rooklin2015alphaspace, katigbak2020alphaspace}. They showed that pocket prior can enhance the binding of generated molecules compared to TargetDiff \cite{guan20233d}. DECOMPOPT \cite{zhou2024decompopt} applies iterative and controllable optimization by conditioning on local substructures applying techniques like R-group optimization and scaffold hopping. For example, the model was used to enhance the drug-likeness metrics such as QED and SAS of the generated molecules while maintaining the binding affinity to the target by conditioning on the molecular arms, making interactions with the bonding pocket \cite{zhou2024decompopt}. \\
Binding Adaptive Diffusion Models (BindDM) \cite{huang2024bindingadaptive} is another model that adaptively generates molecules in reverse diffusion by extracting the essential binding subcomplex graphs at each time step from the protein-ligand complex graph with a learnable structural pooling. The two hierarchies of the complex graph and its subcomplex graph interact during that process through two cross-hierarchy interaction nodes: complex-to-subcomplex (C2S) and subcomplex-to-complex (S2C). BindDM \cite{huang2024bindingadaptive} has a similar performance to TargetDiff \cite{guan20233d} and DecompDiff \cite{guan2024decompdiff}. KGDiff \cite{Qian_Huang_Tu_Xu_2023} also integrates the chemical knowledge of protein-ligand binding affinity to direct the denoising process at each step. Compared to TargetDiff \cite{guan20233d}, it achieved an average Vina Score higher by 46.2\% on the CrossDocked2020 dataset \cite{crossdock}. IPDiff \cite{huang2024proteinligand} applies a similar strategy to BindDM \cite{huang2024bindingadaptive} where a protein-ligand interaction prior network (IPNET) is pre-trained to learn ligand-protein interactions from the chemical properties and 3D structures. Then, the pretrained IPNET serves as a prior to enhance the generative diffusion process. They also introduce two prior strategies: prior-shifting, where the diffusion process is shifted based on protein-molecule interactions learned by IPNET, and prior-conditioning, where the diffusion process is conditioned on previously estimated protein-ligand complexes.  \\
PROMPTDIFF \cite{yang2024promptbased} is also a target-aware generative model that uses a selected set of ligand prompts—that is, compounds with desired characteristics like high binding affinity to the target, drug-likeness, and synthesizability—to direct the generative process towards generating similar molecules meeting the design requirements. The model employs a geometric protein-molecule interaction network (PMINet) that extracts information about the interactions between protein-ligand pairs as embeddings that can be used as a prompt to steer the diffusion process in desirable directions. The model proposes two approaches for promoting the generative process: Self-Promoting, where the embeddings are extracted from the generated ligand at each time step, and Exemplar Prompting, where embeddings extracted from exemplar ligands with the desired properties by pre-trained PMINet are used to guide the reverse generation process \cite{yang2024promptbased}. \\
PMDM \cite{Huang_Xu_Yu_Zhao_Chen_Han_Xie_Li_Zhong_Wong} adopted a novel approach where they conditioned molecular generation on both local and global molecular dynamics trajectories. The model employed an EGNN to model the pocket ligand structures where the pocket geometry served as the condition and a SchNet to generate the conditional protein semantic information encodings and to embed the ligand atom feature into an intermediate representation where the two are later fused using a cross-attention mechanism. \\
The Molecular Out-Of-distribution Diffusion (MOOD) \cite{lee2023exploring} model is a score-based diffusion model that aims to generate molecules with desired chemical properties such as binding affinity, drug-likeness, and synthesizability. The model trains a separate network to predict a particular property, and then the gradients from the property predictor are used to guide the reverse-time diffusion process. This Out-Of-distribution guidance is controlled by a hyperparameter $\lambda \in [0, 1)$ that can be adapted to different magnitudes based on the primary goal of generation. MOOD \cite{lee2023exploring} was able to generate molecules satisfying several constraints at the same time, and that was reflected by the high Novel top $5\%$ docking score (the average DS of the top $5\%$ unique molecules with $\text{QED} > 0.5$ and $\text{SA} < 5$) of the generated compounds for five protein targets. PILOT \cite{cremer2024pilot} also aims to achieve multi-objective conditioning using importance sampling. The model combines pocket conditioning with synthetic accessibility (SA) guidance. First, the model is unconditionally pre-trained on the Enamine Real Diversity subset present in the ZINC database, and then the model is fine-tuned using the pocket-conditioned CrossDocked2020 dataset \cite{crossdock}. Finally, importance sampling is used at inference to guide the diffusion process towards a specific property such as SA. \\
Similar to VoxMol \cite{pinheiro20243d}, a score-based conditional model, VoxBind \cite{pinheiro2024structurebased} was developed to perform target-aware molecular generation using 3D atomic density grids. VoxBind applied a two-step approach where a Langevin MCMC samples noisy ligands, and then clean molecules are predicted using a conditional neural empirical Bayes (NEB) denoiser \cite{saremi2020neural, pinheiro20243d}. The model managed to achieve successful Vina Scores on the n CrossDocked2020 dataset \cite{crossdock}, surpassing DiffSBDD \cite{\cite{schneuing2023structurebased}}, TargetDiff  \cite{guan20233d}, and DecompDif \cite{guan2024decompdiff}. MolSnapper \cite{Ziv_Marsden_Deane_2024} is another recently published model where the diffusion process is guided by 3D pharmacophores incorporating features crucial for making interactions with the target protein such as hydrogen bonds, charge interactions, and lipophilic contacts. The model used Moldiff \cite{peng2023moldiff} for molecule generation with further guidance from the target protein. \\
Moreover, some models were designed to use fragment-based drug design for generating 3D compounds conditioned on protein pockets using fragments such as DiffLinker \cite{igashov2022equivariant}, FragDiff \cite{peng2023pocketspecific} and AutoFragDiff \cite{ghorbani2023autoregressive} which will be discussed in the following section.

\section{Other Applications}
\begin{figure}[ht]
    \centering
    \includegraphics[width=1\linewidth]{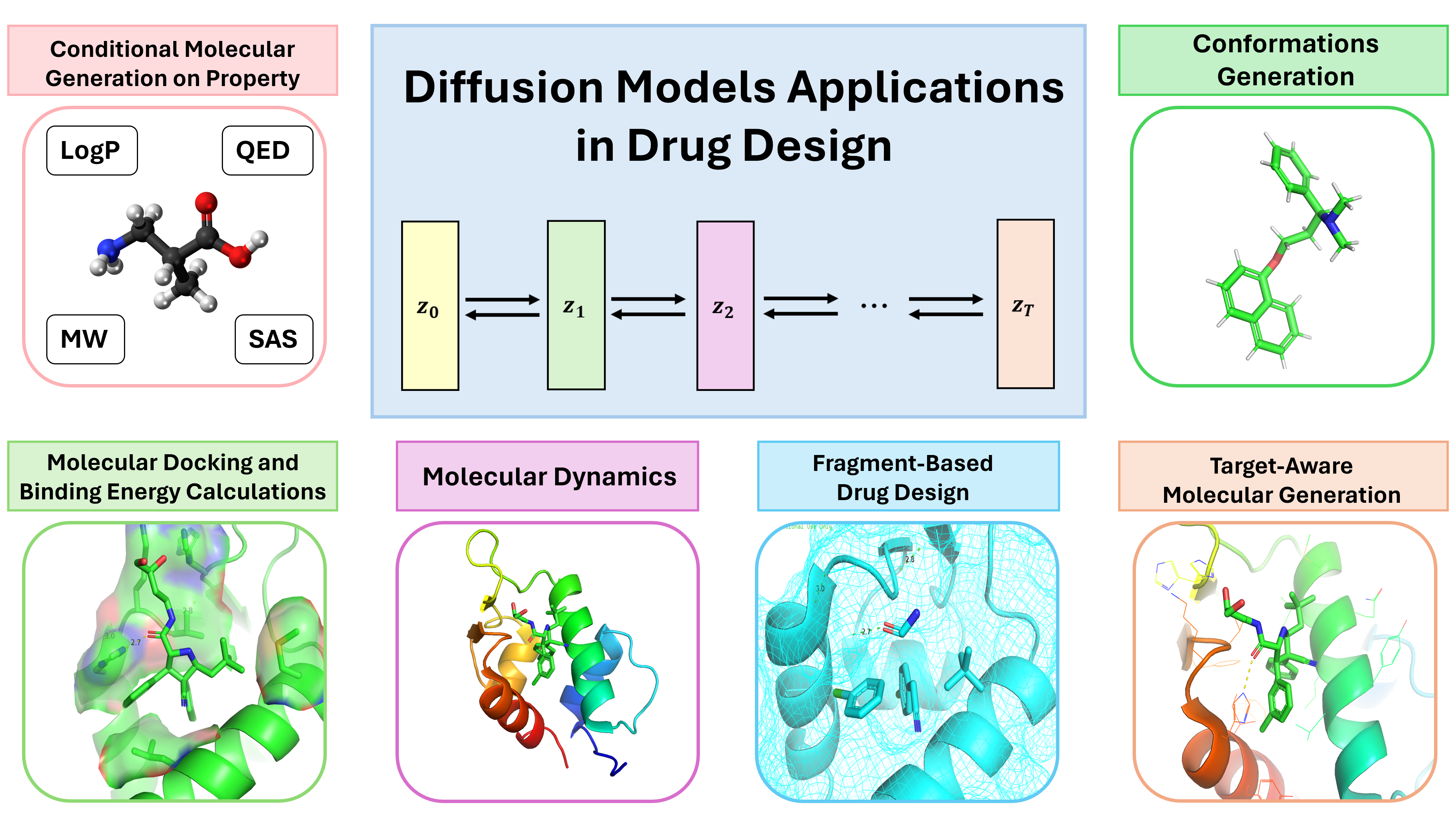}
    \caption{Some applications of diffusion models in  the drug discovery process.}
    \label{fig6}
\end{figure}

In addition to generating molecules in the 3D space, diffusion models have been developed for other applications crucial for the drug design process, such as conformations generation, molecular docking, molecular dynamics, and fragment-based drug design or linker generation (Figure. \ref{fig6} and Table. \ref{applications_table}).

\subsection{Fragment-Based Drug Design and Linker Design}
DiffLinker \cite{igashov2022equivariant} is an EGNNs-based 3D-conditional diffusion model for linker generation. The model applies the concept of fragment-based drug design, where it takes a set of disconnected fragments as a condition and connects them. The linker attachment sites and their size, specifically the required number of atoms, can be predicted automatically by DiffLinker. Similar to the mentioned above SBDD models that applied linker design such as DiffSBDD \cite{schneuing2023structurebased} and PMDM \cite{Huang_Xu_Yu_Zhao_Chen_Han_Xie_Li_Zhong_Wong}, DiffLinker was also used with protein pockets as a condition to the linker design. \\
FragDiff \cite{peng2023pocketspecific} is an autoregressive model that generates 3D molecules conditioned on protein pockets fragment-by-fragment using an E(3)-equivariant diffusion generative model. AutoFragDiff \cite{ghorbani2023autoregressive} follows a similar approach, but it employs geometric vector perceptrons (GVPs) to predict molecular fragments conditioned on molecular scaffolds and protein pockets. \\
Selective iterative latent variable refinement (SILVR) \cite{Runcie_2023} model generated molecules conditioned on fragment hits screened against a protein binding site. Hit fragments first go through the forward diffusion process by adding Gaussian noise at each time step, generating a set of reference fragments at different noise levels for different time steps. Then, the set of noisy fragments is used to condition the EDM during the reverse diffusion process to generate new samples similar to the reference fragments.

\subsection{Conformations Generation}
Conformer generation is a crucial step in drug design, and it involves predicting the various spatial arrangements, or conformations, that a molecule can adopt. Several diffusion models have been designed to efficiently sample molecules' conformational space efficiently. Dynamic Graph Score Matching (DGSM) \cite{Luo2021PredictingMC} is a score-based based that aims to predict equilibrium molecular conformations by modeling both local and long-range interactions using Graph Neural Networks (GNNs). Molecular graphs are constructed using fully connected graphs with a cutoff distance where each node represents an atom within a molecule and is assumed to interact with all the atoms within the sphere constructed by the cutoff distance. Geodiff \cite{xu2022geodiff} employs an equivariant convolutional layer, called graph field network (GFN), to model the molecules where each atom is treated as a particle, and the reverse diffusion process trains a Markov chain to generate stable conformations of the molecules. Geodiff outperformed DGSM \cite{Luo2021PredictingMC} in terms of matching and coverage scores for the GEOM-DRUGs dataset \cite{GEOM}.\\
Torsional diffusion \cite{jing2023torsional} generates conformations using torsional Boltzmann generators by operating only on a hypertorus defined by torsional angles, and a diffusion model is scored using an extrinsic-to-intrinsic model that predicts a torsional score (extrinsic coordinates) unique to a molecule and takes the 3D point cloud representation of its conformer in Euclidean space (intrinsic coordinates) as input. \\
SDEGen \cite{D2SC04429C} aims to model low-energy conformations using a stochastic differential equations (SDE) model that predicts the injected Gaussian noise. Molecular conformations are represented as molecular graphs extended by adding auxiliary bonds such as the two-hop edges (1–3 angle interaction) and three-hop edges (1–4 dihedral angle interaction). Multilayer Perceptrons (MLPs) are used to embed Gaussian noise, interatomic distances matrix, high-dimensional space, and the edge information and add them together, forming distance embedding. Then, Graph Isomorphism Networks (GINs) \cite{hu2020strategies}, a type of GNNs, combine the distance embeddings with atom features embeddings and perform the reverse denoising diffusion process by updating the distance embedding until it gets mapped to the dimensional space of the Gaussian noise. SDEGen reported high coverage and matching scores on the QM9 \cite{QM9} and GEOM-DRUGs \cite{GEOM} datasets compared to other models.

\subsection{Molecular Docking}
Molecular docking aims to predict the optimal binding orientation (pose) of the ligand in the binding pocket of the receptor by simulating the interaction between the two molecules and then scoring the pose based on those interactions. DiffDock \cite{corso2023diffdock} is an SDE score-based generative model (Score SDE) trained to predict ligand poses. Starting from a random pose, the reverse diffusion process operates over translations, rotations, and torsion angles and samples several poses. Another model called a confidence model is trained to score those poses and rank them according to a confidence score that estimates the most likely sample. DiffDock was able to achieve significant accuracy on the PDBBind dataset \cite{doi:10.1021/acs.accounts.6b00491} by retrieving $38.2\%$ of the ligand poses with root-mean-square deviation (RMSD) values $< 2$ Å and $63.2\%$ of the ligands with RMSD $< 5$ Å outperforming the traditional docking methods that require a parameterized physics-based scoring function and a search algorithm such as VINA \cite{10.1002/jcc.21334} as well as deep learning-based methods such as GNINA \cite{McNutt}. DiffDock-PP \cite{ketata2023diffdockpp} is similar to DiffDock \cite{corso2023diffdock}, but it performs rigid protein-protein docking instead. \\
PLANTAIN \cite{brocidiacono2023plantain} is a physics-inspired diffusion model that aims to predict and score ligands' pose within the protein pocket by minimizing a score function. The model used the protein binding pocket encodings, the 2D ligand representation, and ligand-ligand and ligand-residue interatomic distances to iteratively refine a ligand pose and score it. PLANTAIN could predict $21.5\%$ of the ligands' poses within protein pockets with RMSD values $< 2$ Å on the CrossDocked dataset \cite{crossdock}, also outperforming VINA \cite{10.1002/jcc.21334} and GNINA \cite{McNutt}. DiffEE \cite{Nakata_Mori_Tanaka_2023} takes a similar approach where a large-scale protein language model (PLM) encodes input protein sequences, and an iterative process updates the 3D molecular graph and sample ligand poses. DiffEE had comparable results to VINA \cite{10.1002/jcc.21334} and GNINA \cite{McNutt} on the PDBBind benchmark \cite{doi:10.1021/acs.accounts.6b00491} for the $2$ Å and $5$ Å RMSD cutoffs. \\
PocketCFDM \cite{Voitsitskyi_Bdzhola} used data augmentation of binding pockets where several artificial binding pockets are created around the ligand molecule to statistically simulate the non-bond interactions found in actual protein-ligand complexes. PocketCFDM's statistical approach was able to achieve an accuracy of $23.9\%$ with an RMSD cutoff of $2$ Å on the PDBbind dataset \cite{doi:10.1021/acs.accounts.6b00491}, scoring below DiffDock; however, it performed better on the non-bond interactions, steric clash numbers, and inference speed. \\
NeuralPLexer \cite{qiao2023statespecific} is an SDE-based diffusion model that can predict protein-ligand complex structures at atomistic resolution by training a primary network composed of an Equivariant Structure Denoising Module (ESDM) and coarse-grained, auto-regressive contact prediction module (CPM). Protein language model (PLM) features and structure templates are first retrieved from the input protein sequence. Next, using molecular graph representations of the input ligands and the set of PLM and template features as inputs, NeuralPLexer samples a set of binding protein-ligand complex conformations. The model was able to outperform DiffDock in rigid blind protein-ligand docking and predict up to $78\%$ of the ligands' poses within protein pockets with RMSD values $<2$ Å from the reference PDBBind2020 benchmark \cite{doi:10.1021/acs.accounts.6b00491}. \\
Neural Euler's Rotation Equations (NERE) \cite{jin2023unsupervised} is an unsupervised denoising score matching (DSM) energy-based diffusion model that simulates the force and torque between the atoms of the ligand and the protein to predict rotation. The model's log-likelihood is defined as the binding affinity of the ligand-protein complex and trains an SE(3) equivariant rotation prediction network where force is the gradient of the energy function with respect to atom coordinates. NERE was able to achieve Pearson correlation values of 0.656 and 0.651 between true binding affinity and predicted energy on the crystal and docked structures of the PDBBind benchmark, respectively  \cite{doi:10.1021/acs.accounts.6b00491}. \\
The aforementioned diffusion-based docking methods report comparable or better performance than traditional approaches. However, a study by Yu et al. \cite{yu2023deep} argued that deep learning-based models are superior in docking on the whole protein. At the same time, traditional methods are better at docking on given pockets. Therefore, when pocket searching was performed before docking, traditional methods achieved better results for docking in the predetermined pockets than in the entire protein compared to deep learning methods. Moreover, PoseBusters \cite{buttenschoen2023posebusters} found that deep-learning-based tools, including DiffDock \cite{corso2023diffdock}, do not generate physically valid poses. However, those studies only covered DiffDock \cite{corso2023diffdock} and not the following models. With the fast-paced advancement of diffusion-based models, there is a massive potential for further improvement to surpass traditional models in pocket searching and given pocket docking.

\subsection{Molecular Dynamics}
Molecular dynamics (MD) is a computational modeling technique employed to investigate the time-dependent behavior of atoms and molecules by simulating their physical motions through classical mechanics. It has several applications, such as understanding binding mechanisms between drug molecules with their target proteins, protein-protein interactions, and Investigating the dynamics of biomolecules such as proteins and nucleic acid. DiffMD \cite{wu2023diffmd} is a score-based diffusion model aiming to accelerate molecules' MD. DiffMD employs an equivariant geometric transformer to calculate the gradient of the log density of molecular conformations where the directions and velocities of atomic motions are represented by 3D spherical Fourier-Bessel bases. DiffMD outperforms state-of-the-art baselines such as EGNNs \cite{satorras2022en} on MD17 \cite{Chmiela_2017} and isomers of $\text{C}_{7}\text{O}_{2}\text{H}_{10}$ datasets. Another model \cite{arts2023one} aims to estimate a coarse-grained (CG) force field as a Denoising Force Field (DFF) using a score-based diffusion model. The model successfully enhanced performance on various protein simulations for up to 56 amino acid systems. 

\afterpage{\clearpage}

\begin{table}[ht]
  \centering
  \renewcommand{\arraystretch}{1.5}
  \begin{tabularx}{\textwidth}{|p{2.5cm}|p{3cm}p{3.5cm}p{2.5cm}p{4.0cm}|}
    \toprule
    Application    &  Model name & Condition/Guidance & Framework & Network architecture  \\
    \midrule
    \hline
    \multirow{12}{6em}{Target-conditioned molecular generation} &  DiffSBDD \cite{schneuing2023structurebased} & Protein pockets & DDPM & EGNNs \cite{satorras2022en} \\
    & DiffBP \cite{lin2022diffbp} & Protein pockets & DDPM & EGNNs \cite{satorras2022en} \\
    & TargetDiff \cite{guan20233d} & Protein pockets & DDPM & EGNNs \cite{satorras2022en}  \\
    & DECOMPOPT \cite{zhou2024decompopt} & Protein pockets & DDPM & EGNNs \cite{satorras2022en} \\
    & DECOMPDIFF \cite{guan2024decompdiff} & Protein pockets & DDPM & GNNs \\
    & BindDM \cite{huang2024bindingadaptive}  & Protein pockets & DDPM & Hybird 
    \\
    & PROMPTDIFF \cite{yang2024promptbased} & \small{Ligand prompt set and target protein}  & DDPM & PMINet$^a$   \\
    & IPDiff \cite{huang2024proteinligand} & Protein-ligand interaction prior & DDPM & Hybird 
    \\
    & PMDM \cite{Huang_Xu_Yu_Zhao_Chen_Han_Xie_Li_Zhong_Wong} & Protein pockets & SGM & Hybird \\
    & MOOD \cite{lee2023exploring} & Chemical properties & Score SDE & Hybird (GNNs and GCN$^b$)\\
    & PILOT \cite{cremer2024pilot}& Protein pockets and SAS & SGM & GNNs \\
    & KGDiff \cite{Qian_Huang_Tu_Xu_2023} & Protein–ligand binding affinity & DDPM &  GNNs \\
    & VoxBind \cite{pinheiro2024structurebased} & Protein pockets & SGM & 3D U-Net \cite{ronneberger2015unet} \\
    & MolSnapper \cite{Ziv_Marsden_Deane_2024} & 3D pharmacophores and Target protein & DDPM & EGNNs variant \\ 
    \hline
    \multirow{4}{6em}{Fragment-Based Drug Design and Linker Design} & DiffLinker \cite{igashov2022equivariant} &  Protein pockets & DDPM & EGNNs \cite{satorras2022en} \\
    & FragDiff \cite{peng2023pocketspecific} & Protein pockets & DDPM & EGNNs \cite{satorras2022en} \\
    & AutoFragDiff \cite{ghorbani2023autoregressive} & Target protein & DDPM & GVPs$^c$ \\
    &SILVR \cite{Runcie_2023} & Hit fragments & DDPM & EGNNs \cite{satorras2022en} \\
    \hline
    \multirow{4}{6em}{Conformations Generation} & DGSM \cite{Luo2021PredictingMC} & Molecular graph & SGM & GNNs \\
    & Geodiff \cite{xu2022geodiff} & Molecular graph & DDPM & GFN$^d$ \\
    & Torsional diffusion \cite{jing2023torsional} & Molecular graph & Score SDE & Hybird \\
    & SDEGen \cite{D2SC04429C} & Molecular graph & Score SDE & GNNs \\
    \hline
    \multirow{7}{6em}{Molecular Docking} &  DiffDock \cite{corso2023diffdock}  & Protein-ligand complex & Score SDE & GNNs variant (Hybird)  \\
    & DiffDock-PP \cite{ketata2023diffdockpp} & Protein-protein complex & Score SDE & GNNs variant (Hybird)  \\
    & PLANTAIN \cite{brocidiacono2023plantain} & Protein-ligand complex & SGM & GNNs variant (Hybird) \\
    & DiffEE \cite{Nakata_Mori_Tanaka_2023} & Protein sequence and ligand & DDPM & PLM and EGNNs \cite{satorras2022en} \\
    & PocketCFDM \cite{Voitsitskyi_Bdzhola} & Unconditioned  & Score SDE & GNNs variant (Hybird) \\
    & NeuralPLexer \cite{qiao2023statespecific} & Protein-ligand complex & Score SDE& GNNs variant (Hybird)\\
    & NERE \cite{jin2023unsupervised} & Unconditioned & SGM & MPNN$^e$
\cite{yang2019analyzing} \\
     \hline
    \multirow{2}{6em}{Molecular Dynamics} &  DiffMD \cite{wu2023diffmd} & Unconditioned & Score SDE & 
Transformer  \\
    & DFF  \cite{arts2023one} & Unconditioned & SGM & Transformer \\
  \bottomrule
  \hline
  \end{tabularx}
  \caption{Applications of diffusion models in drug design}
  \label{applications_table}
  \footnotesize{$^a$ Protein-molecule interaction network, $^b$ Graph Convolutional Network, $^c$ Geometric Vector Perceptrons, $^d$ Graph Field Network, $^e$ MPNN: Message Passing Neural Network}\\
\end{table}

\section{Evaluation Metrics}
Assessing diffusion models for molecular generation can include various strategies depending on the task. However, the quality of the generated molecules is the most crucial aspect, specifically for the 3D generation, and it requires a multifaceted evaluation strategy. Some evaluation metrics can be used on any molecular generation, whether in 1D, 2D, or 3D, and some metrics are engineered specifically for 3D molecular graphs or drug-like molecules. This section will cover evaluation metrics usually used with molecular generation using diffusion models.

\subsection{Evaluation Metrics for Molecular Generation}
\textbf{Validity - } The percentage of valid molecules, i.e., chemically correct in terms of atoms' valency and consistency of bonds in aromatic rings. This metric is often used with 2D molecular graphs and SMILEs and is evaluated by RDKit. \\
 \newline
\textbf{Novelty - } The percentage of molecules not contained within the training dataset; this metric measures how well the model can generate molecules outside of the dataset in question. \\
\newline
\textbf{Uniqueness - } The percentage of unique and valid molecules in a sample of generated molecules. \\
\newline
\textbf{Atom Stability - } The percentage of atoms with the correct valency (usually used in 3D molecular generation). \\
\newline
\textbf{Molecule Stability - } The percentage of molecules whose atoms are all stable (usually used in 3D molecular generation). \\
\newline
\textbf{Connectivity/Connected - } The percentage of connected molecules, i.e., generated molecules with a single connected component (usually used in molecular graph generation).

\subsection{Similarity Metrics}
Distance metrics are crucial in quantifying the similarity between generated molecules' data distribution and training data distribution in molecular generation tasks. Two widely used metrics for this purpose are the Maximum Mean Discrepancy (MMD) and the Jensen-Shannon (JS) divergence. They are employed to evaluate the distributions of bond lengths, bond angles, and dihedral angles in generated samples compared to the original dataset. \\
\newline
\textbf{Maximum mean discrepancy (MMD) - } A formulation that measures the distance between two probability distributions by embedding probabilities in a reproducing kernel Hilbert space (RKHS) \cite{NIPS2016_5055cbf4}. Given two distributions 
$P$ and $Q$, with corresponding feature maps $\phi(x)$ and $\psi(y)$, the MMD can be expressed as:
\begin{align*}
MMD^2(P, Q) = \left\lVert \mathbb{E}_{x \sim P}[\phi(x)] - \mathbb{E}_{y \sim Q}[\psi(y)] \right\rVert^2_{\mathcal{H}}
\end{align*}
where $\mathbb{E}_{x \sim P}[\phi(x)]$ and $\mathbb{E}_{y \sim Q}[\psi(y)]$ are the expected value of the feature map of samples drawn from distributions $P$, and $Q$ respectively, and $\lVert \cdot \rVert^2_{\mathcal{H}}$ is the norm in the RKHS \cite{NIPS2016_5055cbf4}. \\
\newline
\textbf{Jensen Shannon (JS) divergence - } A metric that quantifies the similarity or dissimilarity between two probability distributions based on the Kullback-Leibler (KL) divergence. Hence, given two distributions $P$ and $Q$, the JS divergence between them should be:
\begin{align*}
    JS(P \parallel Q) = \frac{1}{2} \cdot KL\left(P \parallel M\right) + \frac{1}{2} \cdot KL\left(Q \parallel M\right)
\end{align*}
where $M = \frac{1}{2}(P+Q)$ is the average distribution, and $KL$ is the  Kullback-Leibler divergence and it takes the form:
$$KL(P \parallel Q) = \int_{-\infty}^{\infty} P(x) \cdot \log \left(\frac{P(x)}{Q(x)} \right) dx
$$

\subsection{Drug-Likeness Metrics}
\textbf{logP - the logarithm of the partition coefficient} A measurement of the compound's distribution between two immiscible phases, usually between an organic phase (usually octanol) and an aqueous phase, it is frequently used to assess a compound's lipophilicity, which is an important aspect in determining its absorption, distribution, metabolism, and excretion (ADME) properties. \\
\newline
\textbf{QED - Quantitative Estimate of Druglikeness} An integrative score to assess a compound's likelihood of becoming a drug, it combines a variety of main molecular properties that fall within the range of known drugs. \\
\newline
\textbf{SAS - synthetic accessibility score} Evaluates the synthetic feasibility of the generated molecules, and it is calculated as a sum of fragment scores and complexity penalty \cite{ertl2009estimation}. \\
\newline
\textbf{Lipinski - } Average number of rules satisfied from the Lipinski rule of five, a loose rule for evaluating the drug-likeness of molecules according to five rules the molecule should satisfy: no more than five hydrogen bond donors, no more than ten hydrogen bond acceptors. Molecular mass less than 500 Da, and a logP value less than five \cite{lipinski2012experimental}.

\subsection{Evaluation Metrics for Propriety-Based Conditional Generation}
Conditional generation can be performed with any property. However, it is often used with the QM9 dataset as a proof of concept where a property classifier network, usually an EGNN, is trained on half of the QM9 dataset, while the model is trained on the other half, and then the model is given the property value as an input and is asked to sample molecules with that property value. The mean absolute error (MAE) between the input property values and the predicted values of the generated molecules is used to evaluate the model’s ability of conditional generation. The six OM9 properties often conditioned on are: 
\begin{itemize}
    \item \textbf{$\mathbf{\alpha}$ - Polarizability} in cubic Bohr radius ($\text{Bohr}^3$).
    \item $\mathbf{\mu}$\textbf{ - Dipole moment} in debyes ($D$).
    \item $\mathbf{\epsilon_{HOMO}}$\textbf{ - Highest Occupied Molecular Orbit Energy} in millielectron volts ($meV$).
    \item $\mathbf{\epsilon_{LUMO}}$\textbf{ - Lowest Unoccupied Molecular Orbit Energy} in millielectron volts ($meV$).
    \item $\mathbf{\Delta \epsilon}$\textbf{ - Difference between $\mathbf{\epsilon_{HOMO}}$ and $\mathbf{\epsilon_{LUMO}}$} in millielectron volts ($meV$).
    \item $\mathbf{C_v}$\textbf{ - Molar Heat Capacity} at $298.15 K$ in calories per Kelvin per mole ($\frac{cal}{mol} K$).
\end{itemize}

\subsection{Evaluation Metrics for Structure-Based Generation}
\textbf{Vina Score - } A scoring function that estimates the binding affinity between a ligand (small molecule) and a protein target in molecular docking simulations using Autodock vina \cite{10.1002/jcc.21334}. \\
\newline
\textbf{Vina Min - } The same scoring function as Vina Score, but the Vina platform conducts a local structure minimization before estimation \cite{10.1002/jcc.21334}. \\
\newline
\textbf{Vina Dock - } The same scoring function as Vina Score, but molecules are re-docked before being scored, and it reflects the best possible binding affinity for the generated molecules \cite{10.1002/jcc.21334}. \\
\newline
\textbf{Vina Energy - } An energy estimation of the binding affinity between a ligand (small molecule) and a protein target in molecular docking simulations using Autodock vina \cite{10.1002/jcc.21334}. \\
\newline
\textbf{High-Affinity Percentage - } Percentage of compounds that, upon binding to the target protein, have a lower Vina energy than the reference (ground-truth) molecule. \\
\newline
\textbf{Diversity - } The average pairwise dissimilarity between all generated molecules for each target pocket is calculated as (1 - Tanimoto similarity) between pairs of Morgan fingerprints to measure the diversity of generated molecules for each pocket. \\

\subsection{Evaluation Metrics for Conformations Generation}
Evaluation metrics for evaluating conformers are all based on the root-mean-square deviation (RMSD), which can be calculated as the normalized Frobenius norm of the atoms of two superimposed molecules aligned using the Kabsch algorithm \cite{kabsch1976solution} using the following formula:
\begin{align*}
\text{RMSD}(C, \hat{C}) = \sqrt{\frac{\sum_{i=1}^{N} (x_i - x_i^{ref})^2 + (y_i - y_i^{ref})^2 + (z_i - z_i^{ref})^2}{N}}
\end{align*}
Where  $\hat{C}$ is the predicted molecule conformer with a set of $ i \in \{1, 2, \dots, N \} $ atomic coordinates $(x_i, y_i, z_i)$ and $C$ is the reference molecule conformer with the corresponding reference atomic positions $(x_i^{ref}, y_i^{ref}, z_i^{ref})$, and $N$ is the number of atoms. Then, metrics like Average Minimum RMSD (AMR) or Matching (MAT) and Coverage (COV)can be calculated for Precision (P) and Recall (R) from RMSD values of the generated molecular conformations \cite{xu2022geodiff}. \\
\newline
\textbf{COV - } The percentage of structures in one set covered by another, where covering denotes the RMSD between two conformations falling within a specified threshold $\delta$. Given two sets of conformers, $S_g$ as the generated set and $S_r$ as the reference set, coverage for recall $COV-R$ measures the model's ability to find all the reference conformers (percentage of recalled or predicted reference conformers) while coverage for precision $COV-P$ measures the percentage of the conformers the model generates that are relevant, i.e. can be found in the reference set \cite{xu2022geodiff}. Formally, it can be calculated as: 
\begin{align*}
    \text{COV-R}(S_g, S_r) = \frac{1}{|S_r|} \left| \left \{
    C \in S_r | \text{RMSD}(C, \hat{C}) \leq \delta,    \hat{C} \in S_g 
    \right \}
    \right|
\end{align*} 
\textbf{MAT/AMR - } The average RMSD between conformers in one set and their nearest neighbor in another. i.e., the conformers' average minimum RMSD (AMR), where lower MAT scores reflect the generation of more realistic conformations. Matching for recall (MAT-R) and precision (MAT-P) are defined similarly to the coverage metric \cite{xu2022geodiff}.
\begin{align*}
    \text{MAT-R}(S_g, S_r) = \frac{1}{|S_r|} \sum_{
    C \in S_r} \underset{\hat{C} \in S_g  }{
    \text{min}}   
     \text{RMSD}(C, \hat{C}) 
\end{align*}

\section{Current Limitations}
Diffusion models have significantly advanced molecular generation in general, specifically 3D molecular graph generation. However, there are still some challenges to advance the field further. \\
Chirality is one of the current limitations of the 3D generative models as most of the diffusion models are E(3) equivariant, meaning they are not sensitive to chirality. This issue was only considered when designing the reverse diffusion neural network of the GCDM model \cite{morehead2024geometrycomplete}, GCPNET \cite{morehead2023geometrycomplete}, and similarly, chirality should be considered when designing novel models or improvements over the current state-of-the-art models. \\
A second limitation shared by molecular generation models, in general, is the disconnect between metrics and real-world performance in different applications like drug discovery and material design. While metrics like synthetic accessibility score (SAS) \cite{ertl2009estimation} and Vina docking scores \cite{10.1002/jcc.21334} or binding affinities serve as good estimates, it still does not guarantee that this molecule can be found in libraries of compounds or that it may be synthesized using current methods of synthetic chemistry or the cost efficiency of synthesizing those molecules compared to other possible candidates from other regions of the chemical space. Moreover, the black-box nature of deep generative models in general, including diffusion models, makes it challenging to trace why diffusion models generate specific molecules or conformations. This lack of explainability complicates identifying and addressing potential biases towards certain chemical spaces or limitations in the generation process. \\
A third limitation is related to data availability in chemical applications. Due to the higher cost of experimental validation, such as structure-based applications in drug design and material design applications, the data available for training is still limited and does not cover important aspects such as pharmacokinetics and toxicity properties of the generated molecules. The data can also introduce bias toward known space, limiting the discovery of truly novel and groundbreaking structures, especially when the model is optimized using metrics that reward that bias, such as the Maximum Mean Discrepancy (MMD) distance and other similarity scores. The lack of data is a general problem in several drug design applications, not only in molecular generation. To solve this issue for some applications, Hu et al. used diffusion models to generate synthetic data specific to certain tasks such as pharmacokinetic properties, toxicity, and hERG \cite{du2011hergcentral}. \\
A more specific limitation to the 3D generation is the lack of a unified benchmark; even the widely used metrics, such as 3D stability, are defined differently in each model. For example, some models such as MiDi \cite{vignac2023midi} calculate stability over the adjacency matrix generated by the model for covalent bond types, and other models such as JODO \cite{huang2023learning} calculate the covalent bonds based on distance cutoffs from the generated 3D coordinates and then evaluate the percentage of stable atoms accordingly. However, even simple definitions like what is considered a stable atom are not unified. For example, some models such as MiDi \cite{vignac2023midi} consider ionized atoms stable, so a carbon atom is allowed to have a valency of 4 (formal charge (FC) = 0), 3 (FC = 1, -1), or 5 (FC = -1), and other models such as JODO \cite{huang2023learning} and EDM \cite{hoogeboom2022equivariant} only consider an atom stable if its formal charge is equal to zero. Moreover, conditional generation of molecules on a quantum mechanics' property such as the QM9 polarizability ($\alpha$), and dipole moment ($\mu$), $\dots$, etc., is evaluated based on an EGNN \cite{satorras2022en} predictive model. In contrast, the EGNN model itself has a margin of error from the ground truth (Density-functional theory (DFT) calculations). Given the advancement of computational resources, it would be better to consider generating a reasonable sample of molecules and evaluating the generated molecules using DFT calculations directly. \\
A further limitation of diffusion models is the high training and sampling cost, especially for complex data like molecules. Those computational demands scale significantly with increasing the size of molecules. However, developing new model architectures and training strategies can help achieve better training and sampling from those models. \\
PoseCheck \cite{harris2023benchmarking}, a benchmarking study designed to evaluate deep generative models designed for SBDD, reported another limitation of diffusion models in target-aware generation where the two diffusion-based models (DiffSBDD \cite{schneuing2023structurebased} and TargetDiff \cite{guan20233d}) generate highly-strained molecules with high steric clashes with the protein pocket and suggested penalizing steric clashes to avoid this issue. \\

\section{Conclusions}
Diffusion models hold immense potential for various chemical applications. The early successes motivate intense activity to address challenges in data availability, model performances and relevance to different applications persist. By addressing these limitations and actively researching new approaches and model architectures, diffusion models can be used effectively in various chemical applications. Given the large datasets available and decades of intensive research, drug discovery represents an ideal platform for further development of diffusion models.

\section{Acknowledgements}
I, Amira, thank Sijie Fu, Calvin Gang, Chenghui Zhou, and Euxhen Hasanaj for the helpful discussions and comments on diffusion models that led to this review. Also, my special thanks go to Arav Agarwal for helping with data collection and for his valuable feedback and comments. AI tools, including ChatGPT, Gemini, and Quillbot, were used in limited tasks while writing this article.

\section{Conflict of Interest}
The authors declare no competing interests.

\newpage

\bibliographystyle{unsrt}
\bibliography{main}

\begin{thebibliography}{100}

\bibitem{Pang_Qiao_Zeng_Zou_Wei_2023}
Chao Pang, Jianbo Qiao, Xiangxiang Zeng, Quan Zou, and Leyi Wei.
\newblock Deep generative models inde novodrug molecule generation.
\newblock {\em Journal of Chemical Information and Modeling}, Nov 2023.

\bibitem{10.1093/bib/bbac267}
Wiktoria Wilman, Sonia Wróbel, Weronika Bielska, Piotr Deszynski, Paweł Dudzic, Igor Jaszczyszyn, Jędrzej Kaniewski, Jakub Młokosiewicz, Anahita Rouyan, Tadeusz Satława, Sandeep Kumar, Victor Greiff, and Konrad Krawczyk.
\newblock {Machine-designed biotherapeutics: opportunities, feasibility and advantages of deep learning in computational antibody discovery}.
\newblock {\em Briefings in Bioinformatics}, 23(4):bbac267, 07 2022.

\bibitem{Kim_Mousavi_Yazdi_Zwierzyna_Cardinali_Fox_Peel_Coller_Aggarwal_Maruggi_2024b}
Yoo-Ah Kim, Kambiz Mousavi, Amirali Yazdi, Magda Zwierzyna, Marco Cardinali, Dillion Fox, Thomas Peel, Jeff Coller, Kunal Aggarwal, and Giulietta Maruggi.
\newblock Computational design of mrna vaccines.
\newblock {\em Vaccine}, 42(7):1831–1840, Mar 2024.

\bibitem{Baillif_Cole_McCabe_Bender_2023}
Benoit Baillif, Jason Cole, Patrick McCabe, and Andreas Bender.
\newblock Deep generative models for 3d molecular structure.
\newblock {\em Current Opinion in Structural Biology}, 80:102566, Jun 2023.

\bibitem{doi:10.1517/17460441.2010.497534}
Peter~S Kutchukian and Eugene~I Shakhnovich.
\newblock De novo design: balancing novelty and confined chemical space.
\newblock {\em Expert Opinion on Drug Discovery}, 5(8):789--812, 2010.
\newblock PMID: 22827800.

\bibitem{li2018learning}
Yujia Li, Oriol Vinyals, Chris Dyer, Razvan Pascanu, and Peter Battaglia.
\newblock Learning deep generative models of graphs, 2018.

\bibitem{Li_Zhang_Liu_2018}
Yibo Li, Liangren Zhang, and Zhenming Liu.
\newblock Multi-objective de novo drug design with conditional graph generative model - journal of cheminformatics, Jul 2018.

\bibitem{popova2019molecularrnn}
Mariya Popova, Mykhailo Shvets, Junier Oliva, and Olexandr Isayev.
\newblock Molecularrnn: Generating realistic molecular graphs with optimized properties, 2019.

\bibitem{Elton_Boukouvalas_Fuge_Chung_2019}
Daniel~C. Elton, Zois Boukouvalas, Mark~D. Fuge, and Peter~W. Chung.
\newblock Deep learning for molecular design—a review of the state of the art.
\newblock {\em Molecular Systems Design \&amp; Engineering}, 4(4):828–849, 2019.

\bibitem{simonovsky2018graphvae}
Martin Simonovsky and Nikos Komodakis.
\newblock Graphvae: Towards generation of small graphs using variational autoencoders, 2018.

\bibitem{liu2019constrained}
Qi~Liu, Miltiadis Allamanis, Marc Brockschmidt, and Alexander~L. Gaunt.
\newblock Constrained graph variational autoencoders for molecule design, 2019.

\bibitem{samanta2019nevae}
Bidisha Samanta, Abir De, Gourhari Jana, Pratim~Kumar Chattaraj, Niloy Ganguly, and Manuel Gomez-Rodriguez.
\newblock Nevae: A deep generative model for molecular graphs, 2019.

\bibitem{decao2022molgan}
Nicola~De Cao and Thomas Kipf.
\newblock Molgan: An implicit generative model for small molecular graphs, 2022.

\bibitem{Mercado_Rastemo_Lindelöf_Klambauer_Engkvist_Chen_Bjerrum_2020}
Rocío Mercado, Tobias Rastemo, Edvard Lindelöf, Günter Klambauer, Ola Engkvist, Hongming Chen, and Esben~Jannik Bjerrum.
\newblock Graph networks for molecular design.
\newblock {\em ChemRxiv}, 2020.

\bibitem{madhawa2019graphnvp}
Kaushalya Madhawa, Katushiko Ishiguro, Kosuke Nakago, and Motoki Abe.
\newblock Graphnvp: An invertible flow model for generating molecular graphs, 2019.

\bibitem{shi2020graphaf}
Chence Shi, Minkai Xu, Zhaocheng Zhu, Weinan Zhang, Ming Zhang, and Jian Tang.
\newblock Graphaf: a flow-based autoregressive model for molecular graph generation, 2020.

\bibitem{Zang_2020}
Chengxi Zang and Fei Wang.
\newblock Moflow: An invertible flow model for generating molecular graphs.
\newblock In {\em Proceedings of the 26th ACM SIGKDD International Conference on Knowledge Discovery \&amp; Data Mining}, KDD ’20. ACM, August 2020.

\bibitem{luo2021graphdf}
Youzhi Luo, Keqiang Yan, and Shuiwang Ji.
\newblock Graphdf: A discrete flow model for molecular graph generation, 2021.

\bibitem{Xie_Wang_Li_Lai_Pei_2022}
Weixin Xie, Fanhao Wang, Yibo Li, Luhua Lai, and Jianfeng Pei.
\newblock Advances and challenges in de novo drug design using three-dimensional deep generative models.
\newblock {\em Journal of Chemical Information and Modeling}, 62(10):2269–2279, May 2022.

\bibitem{sohldickstein2015deep}
Jascha Sohl-Dickstein, Eric~A. Weiss, Niru Maheswaranathan, and Surya Ganguli.
\newblock Deep unsupervised learning using nonequilibrium thermodynamics, 2015.

\bibitem{ho2020denoising}
Jonathan Ho, Ajay Jain, and Pieter Abbeel.
\newblock Denoising diffusion probabilistic models, 2020.

\bibitem{kingma2023variational}
Diederik~P. Kingma, Tim Salimans, Ben Poole, and Jonathan Ho.
\newblock Variational diffusion models, 2023.

\bibitem{nichol2021improved}
Alex Nichol and Prafulla Dhariwal.
\newblock Improved denoising diffusion probabilistic models, 2021.

\bibitem{bao2022analyticdpm}
Fan Bao, Chongxuan Li, Jun Zhu, and Bo~Zhang.
\newblock Analytic-dpm: an analytic estimate of the optimal reverse variance in diffusion probabilistic models, 2022.

\bibitem{song2021scorebased}
Yang Song, Jascha Sohl-Dickstein, Diederik~P. Kingma, Abhishek Kumar, Stefano Ermon, and Ben Poole.
\newblock Score-based generative modeling through stochastic differential equations, 2021.

\bibitem{Guo_Liu_Wang_Chen_Wang_Xu_Cheng_2023}
Zhiye Guo, Jian Liu, Yanli Wang, Mengrui Chen, Duolin Wang, Dong Xu, and Jianlin Cheng.
\newblock Diffusion models in bioinformatics and computational biology.
\newblock {\em Nature Reviews Bioengineering}, Oct 2023.

\bibitem{liu2023generative}
Chengyi Liu, Wenqi Fan, Yunqing Liu, Jiatong Li, Hang Li, Hui Liu, Jiliang Tang, and Qing Li.
\newblock Generative diffusion models on graphs: Methods and applications, 2023.

\bibitem{hoogeboom2022equivariant}
Emiel Hoogeboom, Victor~Garcia Satorras, Clément Vignac, and Max Welling.
\newblock Equivariant diffusion for molecule generation in 3d, 2022.

\bibitem{satorras2022en}
Victor~Garcia Satorras, Emiel Hoogeboom, and Max Welling.
\newblock E(n) equivariant graph neural networks, 2022.

\bibitem{https://doi.org/10.13140/rg.2.2.26493.64480}
{Mengchun Zhang}, Maryam Qamar, {Taegoo Kang}, Yuna Jung, {Chenshuang Zhang}, Sung-Ho Bae, and {Chaoning Zhang}.
\newblock A survey on graph diffusion models: Generative ai in science for molecule, protein and material.
\newblock 2023.

\bibitem{yang2024diffusion}
Ling Yang, Zhilong Zhang, Yang Song, Shenda Hong, Runsheng Xu, Yue Zhao, Wentao Zhang, Bin Cui, and Ming-Hsuan Yang.
\newblock Diffusion models: A comprehensive survey of methods and applications, 2024.

\bibitem{vignac2023midi}
Clement Vignac, Nagham Osman, Laura Toni, and Pascal Frossard.
\newblock Midi: Mixed graph and 3d denoising diffusion for molecule generation, 2023.

\bibitem{morehead2024geometrycomplete}
Alex Morehead and Jianlin Cheng.
\newblock Geometry-complete diffusion for 3d molecule generation and optimization, 2024.

\bibitem{xu2022geodiff}
Minkai Xu, Lantao Yu, Yang Song, Chence Shi, Stefano Ermon, and Jian Tang.
\newblock Geodiff: a geometric diffusion model for molecular conformation generation, 2022.

\bibitem{vignac2023digress}
Clement Vignac, Igor Krawczuk, Antoine Siraudin, Bohan Wang, Volkan Cevher, and Pascal Frossard.
\newblock Digress: Discrete denoising diffusion for graph generation, 2023.

\bibitem{qiang2023coarsetofine}
Bo~Qiang, Yuxuan Song, Minkai Xu, Jingjing Gong, Bowen Gao, Hao Zhou, Weiying Ma, and Yanyan Lan.
\newblock Coarse-to-fine: a hierarchical diffusion model for molecule generation in 3d, 2023.

\bibitem{song2019sliced}
Yang Song, Sahaj Garg, Jiaxin Shi, and Stefano Ermon.
\newblock Sliced score matching: A scalable approach to density and score estimation, 2019.

\bibitem{song2020generative}
Yang Song and Stefano Ermon.
\newblock Generative modeling by estimating gradients of the data distribution, 2020.

\bibitem{anderson1982reverse}
Brian~DO Anderson.
\newblock Reverse-time diffusion equation models.
\newblock {\em Stochastic Processes and their Applications}, 12(3):313--326, 1982.

\bibitem{karras2022elucidating}
Tero Karras, Miika Aittala, Timo Aila, and Samuli Laine.
\newblock Elucidating the design space of diffusion-based generative models, 2022.

\bibitem{song2022denoising}
Jiaming Song, Chenlin Meng, and Stefano Ermon.
\newblock Denoising diffusion implicit models, 2022.

\bibitem{lu2022dpmsolver}
Cheng Lu, Yuhao Zhou, Fan Bao, Jianfei Chen, Chongxuan Li, and Jun Zhu.
\newblock Dpm-solver: A fast ode solver for diffusion probabilistic model sampling in around 10 steps, 2022.

\bibitem{wu2022diffusionbased}
Lemeng Wu, Chengyue Gong, Xingchao Liu, Mao Ye, and Qiang Liu.
\newblock Diffusion-based molecule generation with informative prior bridges, 2022.

\bibitem{jo2022scorebased}
Jaehyeong Jo, Seul Lee, and Sung~Ju Hwang.
\newblock Score-based generative modeling of graphs via the system of stochastic differential equations, 2022.

\bibitem{huang2023learning}
Han Huang, Leilei Sun, Bowen Du, and Weifeng Lv.
\newblock Learning joint 2d \& 3d diffusion models for complete molecule generation, 2023.

\bibitem{lee2023exploring}
Seul Lee, Jaehyeong Jo, and Sung~Ju Hwang.
\newblock Exploring chemical space with score-based out-of-distribution generation, 2023.

\bibitem{PENG2024122949}
Xinmiao Peng and Fei Zhu.
\newblock Hitting stride by degrees: Fine grained molecular generation via diffusion model.
\newblock {\em Expert Systems with Applications}, 244:122949, 2024.

\bibitem{QM9}
Raghunathan Ramakrishnan, Pavlo~O. Dral, Matthias Rupp, and O.~Anatole von Lilienfeld.
\newblock Quantum chemistry structures and properties of 134 kilo molecules.
\newblock {\em Scientific Data}, 1, 2014.

\bibitem{GEOM}
Simon Axelrod and Rafael Gomez-Bombarelli.
\newblock Geom: Energy-annotated molecular conformations for property prediction and molecular generation, 2022.

\bibitem{ZINC}
John~J. Irwin, Teague Sterling, Michael~M. Mysinger, Erin~S. Bolstad, and Ryan~G. Coleman.
\newblock Zinc: A free tool to discover chemistry for biology.
\newblock {\em Journal of Chemical Information and Modeling}, 52(7):1757–1768, Jun 2012.

\bibitem{MOSES}
Daniil Polykovskiy, Alexander Zhebrak, Benjamin Sanchez-Lengeling, Sergey Golovanov, Oktai Tatanov, Stanislav Belyaev, Rauf Kurbanov, Aleksey Artamonov, Vladimir Aladinskiy, Mark Veselov, and et~al.
\newblock Molecular sets (moses): A benchmarking platform for molecular generation models.
\newblock {\em Frontiers in Pharmacology}, 11, Dec 2020.

\bibitem{crossdock}
Paul~G. Francoeur, Tomohide Masuda, Jocelyn Sunseri, Andrew Jia, Richard~B. Iovanisci, Ian Snyder, and David~R. Koes.
\newblock Three-dimensional convolutional neural networks and a cross-docked data set for structure-based drug design.
\newblock {\em Journal of Chemical Information and Modeling}, 60(9):4200–4215, Aug 2020.

\bibitem{doi:10.1021/acs.accounts.6b00491}
Zhihai Liu, Minyi Su, Li~Han, Jie Liu, Qifan Yang, Yan Li, and Renxiao Wang.
\newblock Forging the basis for developing protein–ligand interaction scoring functions.
\newblock {\em Accounts of Chemical Research}, 50(2):302--309, 2017.
\newblock PMID: 28182403.

\bibitem{adams2021learning}
Keir Adams, Lagnajit Pattanaik, and Connor~W. Coley.
\newblock Learning 3d representations of molecular chirality with invariance to bond rotations, 2021.

\bibitem{huang2022graphgdp}
Han Huang, Leilei Sun, Bowen Du, Yanjie Fu, and Weifeng Lv.
\newblock Graphgdp: Generative diffusion processes for permutation invariant graph generation, 2022.

\bibitem{peng2023moldiff}
Xingang Peng, Jiaqi Guan, Qiang Liu, and Jianzhu Ma.
\newblock Moldiff: Addressing the atom-bond inconsistency problem in 3d molecule diffusion generation, 2023.

\bibitem{xu2024geometricfacilitated}
Can Xu, Haosen Wang, Weigang Wang, Pengfei Zheng, and Hongyang Chen.
\newblock Geometric-facilitated denoising diffusion model for 3d molecule generation, 2024.

\bibitem{dieleman2022continuous}
Sander Dieleman, Laurent Sartran, Arman Roshannai, Nikolay Savinov, Yaroslav Ganin, Pierre~H. Richemond, Arnaud Doucet, Robin Strudel, Chris Dyer, Conor Durkan, Curtis Hawthorne, Rémi Leblond, Will Grathwohl, and Jonas Adler.
\newblock Continuous diffusion for categorical data, 2022.

\bibitem{chen2023analog}
Ting Chen, Ruixiang Zhang, and Geoffrey Hinton.
\newblock Analog bits: Generating discrete data using diffusion models with self-conditioning, 2023.

\bibitem{haefeli2023diffusion}
Kilian~Konstantin Haefeli, Karolis Martinkus, Nathanaël Perraudin, and Roger Wattenhofer.
\newblock Diffusion models for graphs benefit from discrete state spaces, 2023.

\bibitem{xu2023geometric}
Minkai Xu, Alexander Powers, Ron Dror, Stefano Ermon, and Jure Leskovec.
\newblock Geometric latent diffusion models for 3d molecule generation, 2023.

\bibitem{zhu20243mdiffusion}
Huaisheng Zhu, Teng Xiao, and Vasant~G Honavar.
\newblock 3m-diffusion: Latent multi-modal diffusion for text-guided generation of molecular graphs, 2024.

\bibitem{zhou2021graph}
Jie Zhou, Ganqu Cui, Shengding Hu, Zhengyan Zhang, Cheng Yang, Zhiyuan Liu, Lifeng Wang, Changcheng Li, and Maosong Sun.
\newblock Graph neural networks: A review of methods and applications, 2021.

\bibitem{morehead2023geometrycomplete}
Alex Morehead and Jianlin Cheng.
\newblock Geometry-complete perceptron networks for 3d molecular graphs, 2023.

\bibitem{chen2023shapeconditioned}
Ziqi Chen, Bo~Peng, Srinivasan Parthasarathy, and Xia Ning.
\newblock Shape-conditioned 3d molecule generation via equivariant diffusion models, 2023.

\bibitem{hu2020strategies}
Weihua Hu, Bowen Liu, Joseph Gomes, Marinka Zitnik, Percy Liang, Vijay Pande, and Jure Leskovec.
\newblock Strategies for pre-training graph neural networks, 2020.

\bibitem{bao2023equivariant}
Fan Bao, Min Zhao, Zhongkai Hao, Peiyao Li, Chongxuan Li, and Jun Zhu.
\newblock Equivariant energy-guided sde for inverse molecular design, 2023.

\bibitem{ponder2003force}
Jay~W Ponder and David~A Case.
\newblock Force fields for protein simulations.
\newblock {\em Advances in Protein Chemistry}, 66:27--85, 2003.

\bibitem{batatia2023mace}
Ilyes Batatia, Dávid~Péter Kovács, Gregor N.~C. Simm, Christoph Ortner, and Gábor Csányi.
\newblock Mace: Higher order equivariant message passing neural networks for fast and accurate force fields, 2023.

\bibitem{Eastman_Behara_Dotson_Galvelis_Herr_Horton_Mao_Chodera_Pritchard_Wang_et}
Peter Eastman, Pavan~Kumar Behara, David~L. Dotson, Raimondas Galvelis, John~E. Herr, Josh~T. Horton, Yuezhi Mao, John~D. Chodera, Benjamin~P. Pritchard, Yuanqing Wang, and et~al.
\newblock Spice, a dataset of drug-like molecules and peptides for training machine learning potentials.
\newblock {\em Scientific Data}, 10(1), Jan 2023.

\bibitem{elijošius2024zero}
Rokas Elijošius, Fabian Zills, Ilyes Batatia, Sam~Walton Norwood, Dávid~Péter Kovács, Christian Holm, and Gábor Csányi.
\newblock Zero shot molecular generation via similarity kernels, 2024.

\bibitem{doi:10.1021/acs.jctc.4c00232}
Hongni Jin and Kenneth M.~Jr. Merz.
\newblock Liganddiff: de novo ligand design for 3d transition metal complexes with diffusion models.
\newblock {\em Journal of Chemical Theory and Computation}, 0(0):null, 0.
\newblock PMID: 38743854.

\bibitem{oshea2015introduction}
Keiron O'Shea and Ryan Nash.
\newblock An introduction to convolutional neural networks, 2015.

\bibitem{schütt2017schnet}
Kristof~T. Schütt, Pieter-Jan Kindermans, Huziel~E. Sauceda, Stefan Chmiela, Alexandre Tkatchenko, and Klaus-Robert Müller.
\newblock Schnet: A continuous-filter convolutional neural network for modeling quantum interactions, 2017.

\bibitem{huang2022mdm}
Lei Huang, Hengtong Zhang, Tingyang Xu, and Ka-Chun Wong.
\newblock Mdm: Molecular diffusion model for 3d molecule generation, 2022.

\bibitem{pinheiro20243d}
Pedro~O. Pinheiro, Joshua Rackers, Joseph Kleinhenz, Michael Maser, Omar Mahmood, Andrew~Martin Watkins, Stephen Ra, Vishnu Sresht, and Saeed Saremi.
\newblock 3d molecule generation by denoising voxel grids, 2024.

\bibitem{ronneberger2015unet}
Olaf Ronneberger, Philipp Fischer, and Thomas Brox.
\newblock U-net: Convolutional networks for biomedical image segmentation, 2015.

\bibitem{vaswani2017attention}
Ashish Vaswani, Noam Shazeer, Niki Parmar, Jakob Uszkoreit, Llion Jones, Aidan~N. Gomez, Łukasz Kaiser, and Illia Polosukhin.
\newblock Attention is all you need.
\newblock In {\em Advances in Neural Information Processing Systems}, pages 6000--6010. Curran Associates, Inc., 2017.

\bibitem{Guo_Wang_Yu_McKenna_Law_2022}
Yunhui Guo, Chaofeng Wang, Stella~X. Yu, Frank McKenna, and Kincho~H. Law.
\newblock Adaln: A vision transformer for multidomain learning and predisaster building information extraction from images.
\newblock {\em Journal of Computing in Civil Engineering}, 36(5), Sep 2022.

\bibitem{hua2024mudiff}
Chenqing Hua, Sitao Luan, Minkai Xu, Rex Ying, Jie Fu, Stefano Ermon, and Doina Precup.
\newblock Mudiff: Unified diffusion for complete molecule generation, 2024.

\bibitem{dwivedi2021generalization}
Vijay~Prakash Dwivedi and Xavier Bresson.
\newblock A generalization of transformer networks to graphs, 2021.

\bibitem{perez2017film}
Ethan Perez, Florian Strub, Harm de~Vries, Vincent Dumoulin, and Aaron Courville.
\newblock Film: Visual reasoning with a general conditioning layer, 2017.

\bibitem{corso2020principal}
Gabriele Corso, Luca Cavalleri, Dominique Beaini, Pietro Liò, and Petar Veličković.
\newblock Principal neighbourhood aggregation for graph nets, 2020.

\bibitem{huang2023conditional}
Han Huang, Leilei Sun, Bowen Du, and Weifeng Lv.
\newblock Conditional diffusion based on discrete graph structures for molecular graph generation, 2023.

\bibitem{Lin_Xu_Chen_2024}
Jie Lin, Mingyuan Xu, and Hongming Chen.
\newblock Diff-shape: A novel constrained diffusion model for shape based de novo drug design.
\newblock {\em ChemRxiv}, 2024.

\bibitem{zhang2023adding}
Lvmin Zhang, Anyi Rao, and Maneesh Agrawala.
\newblock Adding conditional control to text-to-image diffusion models, 2023.

\bibitem{Arunachalam_Gugler_Taylor_Duan_Nandy_Janet_Meyer_Oldenstaedt_Chu_Kulik_2022}
Naveen Arunachalam, Stefan Gugler, Michael~G. Taylor, Chenru Duan, Aditya Nandy, Jon~Paul Janet, Ralf Meyer, Jonas Oldenstaedt, Daniel~B. Chu, and Heather~J. Kulik.
\newblock Ligand additivity relationships enable efficient exploration of transition metal chemical space.
\newblock {\em The Journal of Chemical Physics}, 157(18), Nov 2022.

\bibitem{Groom_Bruno_Lightfoot_Ward_2016}
Colin~R. Groom, Ian~J. Bruno, Matthew~P. Lightfoot, and Suzanna~C. Ward.
\newblock The cambridge structural database.
\newblock {\em Acta Crystallographica Section B Structural Science, Crystal Engineering and Materials}, 72(2):171–179, Apr 2016.

\bibitem{Brown_2019}
Nathan Brown, Marco Fiscato, Marwin~H.S. Segler, and Alain~C. Vaucher.
\newblock Guacamol: Benchmarking models for de novo molecular design.
\newblock {\em Journal of Chemical Information and Modeling}, 59(3):1096–1108, March 2019.

\bibitem{Hastings_Owen_Dekker_Ennis_Kale_Muthukrishnan_Turner_Swainston_Mendes_Steinbeck_2015}
Janna Hastings, Gareth Owen, Adriano Dekker, Marcus Ennis, Namrata Kale, Venkatesh Muthukrishnan, Steve Turner, Neil Swainston, Pedro Mendes, and Christoph Steinbeck.
\newblock Chebi in 2016: Improved services and an expanding collection of metabolites.
\newblock {\em Nucleic Acids Research}, 44(D1), Oct 2015.

\bibitem{liu-etal-2023-molca}
Zhiyuan Liu, Sihang Li, Yanchen Luo, Hao Fei, Yixin Cao, Kenji Kawaguchi, Xiang Wang, and Tat-Seng Chua.
\newblock {M}ol{CA}: Molecular graph-language modeling with cross-modal projector and uni-modal adapter.
\newblock In Houda Bouamor, Juan Pino, and Kalika Bali, editors, {\em Proceedings of the 2023 Conference on Empirical Methods in Natural Language Processing}, pages 15623--15638, Singapore, December 2023. Association for Computational Linguistics.

\bibitem{Zeng_Yao_Liu_Sun_2022}
Zheni Zeng, Yuan Yao, Zhiyuan Liu, and Maosong Sun.
\newblock A deep-learning system bridging molecule structure and biomedical text with comprehension comparable to human professionals.
\newblock {\em Nature Communications}, 13(1), Feb 2022.

\bibitem{su2022molecular}
Bing Su, Dazhao Du, Zhao Yang, Yujie Zhou, Jiangmeng Li, Anyi Rao, Hao Sun, Zhiwu Lu, and Ji-Rong Wen.
\newblock A molecular multimodal foundation model associating molecule graphs with natural language, 2022.

\bibitem{luo2022fast}
Tianze Luo, Zhanfeng Mo, and Sinno~Jialin Pan.
\newblock Fast graph generation via spectral diffusion, 2022.

\bibitem{schneuing2023structurebased}
Arne Schneuing, Yuanqi Du, Charles Harris, Arian Jamasb, Ilia Igashov, Weitao Du, Tom Blundell, Pietro Lió, Carla Gomes, Max Welling, Michael Bronstein, and Bruno Correia.
\newblock Structure-based drug design with equivariant diffusion models, 2023.

\bibitem{lin2022diffbp}
Haitao Lin, Yufei Huang, Meng Liu, Xuanjing Li, Shuiwang Ji, and Stan~Z. Li.
\newblock Diffbp: Generative diffusion of 3d molecules for target protein binding, 2022.

\bibitem{guan20233d}
Jiaqi Guan, Wesley~Wei Qian, Xingang Peng, Yufeng Su, Jian Peng, and Jianzhu Ma.
\newblock 3d equivariant diffusion for target-aware molecule generation and affinity prediction, 2023.

\bibitem{10.1002/jcc.21334}
Oleg Trott and Arthur~J. Olson.
\newblock Autodock vina: improving the speed and accuracy of docking with a new scoring function, efficient optimization, and multithreading.
\newblock {\em Journal of Computational Chemistry}, 31(2):455--461, 2010.

\bibitem{peng2022pocket2mol}
Xingang Peng, Shitong Luo, Jiaqi Guan, Qi~Xie, Jian Peng, and Jianzhu Ma.
\newblock Pocket2mol: Efficient molecular sampling based on 3d protein pockets, 2022.

\bibitem{liu2022generating}
Meng Liu, Youzhi Luo, Kanji Uchino, Koji Maruhashi, and Shuiwang Ji.
\newblock Generating 3d molecules for target protein binding, 2022.

\bibitem{zhou2024decompopt}
Xiangxin Zhou, Xiwei Cheng, Yuwei Yang, Yu~Bao, Liang Wang, and Quanquan Gu.
\newblock Decompopt: Controllable and decomposed diffusion models for structure-based molecular optimization.
\newblock In {\em The Twelfth International Conference on Learning Representations}, 2024.

\bibitem{guan2024decompdiff}
Jiaqi Guan, Xiangxin Zhou, Yuwei Yang, Yu~Bao, Jian Peng, Jianzhu Ma, Qiang Liu, Liang Wang, and Quanquan Gu.
\newblock Decompdiff: Diffusion models with decomposed priors for structure-based drug design, 2024.

\bibitem{rooklin2015alphaspace}
David Rooklin, Cheng Wang, Joseph Katigbak, Paramjit~S Arora, and Yingkai Zhang.
\newblock Alphaspace: fragment-centric topographical mapping to target protein--protein interaction interfaces.
\newblock {\em Journal of chemical information and modeling}, 55(8):1585--1599, 2015.

\bibitem{katigbak2020alphaspace}
Joseph Katigbak, Haotian Li, David Rooklin, and Yingkai Zhang.
\newblock Alphaspace 2.0: representing concave biomolecular surfaces using $\beta$-clusters.
\newblock {\em Journal of chemical information and modeling}, 60(3):1494--1508, 2020.

\bibitem{huang2024bindingadaptive}
Zhilin Huang, Ling Yang, Zaixi Zhang, Xiangxin Zhou, Yu~Bao, Xiawu Zheng, Yuwei Yang, Yu~Wang, and Wenming Yang.
\newblock Binding-adaptive diffusion models for structure-based drug design, 2024.

\bibitem{Qian_Huang_Tu_Xu_2023}
Hao Qian, Wenjing Huang, Shikui Tu, and Lei Xu.
\newblock Kgdiff: Towards explainable target-aware molecule generation with knowledge guidance.
\newblock {\em Briefings in Bioinformatics}, 25(1), Nov 2023.

\bibitem{huang2024proteinligand}
Zhilin Huang, Ling Yang, Xiangxin Zhou, Zhilong Zhang, Wentao Zhang, Xiawu Zheng, Jie Chen, Yu~Wang, Bin CUI, and Wenming Yang.
\newblock Protein-ligand interaction prior for binding-aware 3d molecule diffusion models.
\newblock In {\em The Twelfth International Conference on Learning Representations}, 2024.

\bibitem{yang2024promptbased}
Ling Yang, Zhilin Huang, Xiangxin Zhou, Minkai Xu, Wentao Zhang, Yu~Wang, Xiawu Zheng, Wenming Yang, Ron~O. Dror, Shenda Hong, and Bin CUI.
\newblock Prompt-based 3d molecular diffusion models for structure-based drug design, 2024.

\bibitem{Huang_Xu_Yu_Zhao_Chen_Han_Xie_Li_Zhong_Wong}
Lei Huang, Tingyang Xu, Yang Yu, Peilin Zhao, Xingjian Chen, Jing Han, Zhi Xie, Hailong Li, Wenge Zhong, Ka-Chun Wong, and et~al.
\newblock A dual diffusion model enables 3d molecule generation and lead optimization based on target pockets.
\newblock {\em Nature Communications}, 15(1), Mar 2024.

\bibitem{cremer2024pilot}
Julian Cremer, Tuan Le, Frank Noé, Djork-Arné Clevert, and Kristof~T. Schütt.
\newblock Pilot: Equivariant diffusion for pocket conditioned de novo ligand generation with multi-objective guidance via importance sampling, 2024.

\bibitem{pinheiro2024structurebased}
Pedro~O. Pinheiro, Arian Jamasb, Omar Mahmood, Vishnu Sresht, and Saeed Saremi.
\newblock Structure-based drug design by denoising voxel grids, 2024.

\bibitem{saremi2020neural}
Saeed Saremi and Aapo Hyvarinen.
\newblock Neural empirical bayes, 2020.

\bibitem{Ziv_Marsden_Deane_2024}
Yael Ziv, Brian Marsden, and Charlotte~M. Deane.
\newblock Molsnapper: Conditioning diffusion for structure based drug design.
\newblock Mar 2024.

\bibitem{igashov2022equivariant}
Ilia Igashov, Hannes Stärk, Clément Vignac, Victor~Garcia Satorras, Pascal Frossard, Max Welling, Michael Bronstein, and Bruno Correia.
\newblock Equivariant 3d-conditional diffusion models for molecular linker design, 2022.

\bibitem{peng2023pocketspecific}
Xingang Peng, Jiaqi Guan, Jian Peng, and Jianzhu Ma.
\newblock Pocket-specific 3d molecule generation by fragment-based autoregressive diffusion models, 2023.

\bibitem{ghorbani2023autoregressive}
Mahdi Ghorbani, Leo Gendelev, Paul Beroza, and Michael~J. Keiser.
\newblock Autoregressive fragment-based diffusion for pocket-aware ligand design, 2023.

\bibitem{Runcie_2023}
Nicholas~T. Runcie and Antonia~S.J.S. Mey.
\newblock Silvr: Guided diffusion for molecule generation.
\newblock {\em Journal of Chemical Information and Modeling}, 63(19):5996–6005, September 2023.

\bibitem{Luo2021PredictingMC}
Shitong Luo, Chence Shi, Minkai Xu, and Jian Tang.
\newblock Predicting molecular conformation via dynamic graph score matching.
\newblock In {\em Neural Information Processing Systems}, 2021.

\bibitem{jing2023torsional}
Bowen Jing, Gabriele Corso, Jeffrey Chang, Regina Barzilay, and Tommi Jaakkola.
\newblock Torsional diffusion for molecular conformer generation, 2023.

\bibitem{D2SC04429C}
Haotian Zhang, Shengming Li, Jintu Zhang, Zhe Wang, Jike Wang, Dejun Jiang, Zhiwen Bian, Yixue Zhang, Yafeng Deng, Jianfei Song, Yu~Kang, and Tingjun Hou.
\newblock Sdegen: learning to evolve molecular conformations from thermodynamic noise for conformation generation.
\newblock {\em Chem. Sci.}, 14:1557--1568, 2023.

\bibitem{corso2023diffdock}
Gabriele Corso, Hannes Stärk, Bowen Jing, Regina Barzilay, and Tommi Jaakkola.
\newblock Diffdock: Diffusion steps, twists, and turns for molecular docking, 2023.

\bibitem{McNutt}
Andrew~T. McNutt, Paul Francoeur, Rishal Aggarwal, Tomohide Masuda, Rocco Meli, Matthew Ragoza, Jocelyn Sunseri, and David~Ryan Koes.
\newblock Gnina 1.0: Molecular docking with deep learning.
\newblock {\em Journal of Cheminformatics}, 13(1), Jun 2021.

\bibitem{ketata2023diffdockpp}
Mohamed~Amine Ketata, Cedrik Laue, Ruslan Mammadov, Hannes Stärk, Menghua Wu, Gabriele Corso, Céline Marquet, Regina Barzilay, and Tommi~S. Jaakkola.
\newblock Diffdock-pp: Rigid protein-protein docking with diffusion models, 2023.

\bibitem{brocidiacono2023plantain}
Michael Brocidiacono, Konstantin~I. Popov, David~Ryan Koes, and Alexander Tropsha.
\newblock Plantain: Diffusion-inspired pose score minimization for fast and accurate molecular docking, 2023.

\bibitem{Nakata_Mori_Tanaka_2023}
Shuya Nakata, Yoshiharu Mori, and Shigenori Tanaka.
\newblock End-to-end protein–ligand complex structure generation with diffusion-based generative models.
\newblock {\em BMC Bioinformatics}, 24(1), Jun 2023.

\bibitem{Voitsitskyi_Bdzhola}
Taras Voitsitskyi, Volodymyr Bdzhola, Roman Stratiichuk, Ihor Koleiev, Zakhar Ostrovsky, Volodymyr Vozniak, Ivan Khropachov, Pavlo Henitsoi, Leonid Popryho, Roman Zhytar, and et~al.
\newblock Augmenting a training dataset of the generative diffusion model for molecular docking with artificial binding pockets.
\newblock {\em RSC Advances}, 14(2):1341–1353, 2024.

\bibitem{qiao2023statespecific}
Zhuoran Qiao, Weili Nie, Arash Vahdat, Thomas F. Miller~III au2, and Anima Anandkumar.
\newblock State-specific protein-ligand complex structure prediction with a multi-scale deep generative model, 2023.

\bibitem{jin2023unsupervised}
Wengong Jin, Siranush Sarkizova, Xun Chen, Nir Hacohen, and Caroline Uhler.
\newblock Unsupervised protein-ligand binding energy prediction via neural euler's rotation equation, 2023.

\bibitem{yu2023deep}
Yuejiang Yu, Shuqi Lu, Zhifeng Gao, Hang Zheng, and Guolin Ke.
\newblock Do deep learning models really outperform traditional approaches in molecular docking?, 2023.

\bibitem{buttenschoen2023posebusters}
Martin Buttenschoen, Garrett~M. Morris, and Charlotte~M. Deane.
\newblock Posebusters: Ai-based docking methods fail to generate physically valid poses or generalise to novel sequences, 2023.

\bibitem{wu2023diffmd}
Fang Wu and Stan~Z. Li.
\newblock Diffmd: A geometric diffusion model for molecular dynamics simulations, 2023.

\bibitem{Chmiela_2017}
Stefan Chmiela, Alexandre Tkatchenko, Huziel~E. Sauceda, Igor Poltavsky, Kristof~T. Schütt, and Klaus-Robert Müller.
\newblock Machine learning of accurate energy-conserving molecular force fields.
\newblock {\em Science Advances}, 3(5), May 2017.

\bibitem{arts2023one}
Marloes Arts, Victor~Garcia Satorras, Chin-Wei Huang, Daniel Zuegner, Marco Federici, Cecilia Clementi, Frank Noé, Robert Pinsler, and Rianne van~den Berg.
\newblock Two for one: Diffusion models and force fields for coarse-grained molecular dynamics, 2023.

\bibitem{yang2019analyzing}
Kevin Yang, Kyle Swanson, Wengong Jin, Connor Coley, Philipp Eiden, Hua Gao, Angel Guzman-Perez, Timothy Hopper, Brian Kelley, Miriam Mathea, Andrew Palmer, Volker Settels, Tommi Jaakkola, Klavs Jensen, and Regina Barzilay.
\newblock Analyzing learned molecular representations for property prediction, 2019.

\bibitem{NIPS2016_5055cbf4}
Ilya~O Tolstikhin, Bharath~K. Sriperumbudur, and Bernhard Sch\"{o}lkopf.
\newblock Minimax estimation of maximum mean discrepancy with radial kernels.
\newblock In D.~Lee, M.~Sugiyama, U.~Luxburg, I.~Guyon, and R.~Garnett, editors, {\em Advances in Neural Information Processing Systems}, volume~29. Curran Associates, Inc., 2016.

\bibitem{ertl2009estimation}
Peter Ertl and Ansgar Schuffenhauer.
\newblock Estimation of synthetic accessibility score of drug-like molecules based on molecular complexity and fragment contributions.
\newblock {\em Journal of Cheminformatics}, 1:1--11, 2009.

\bibitem{lipinski2012experimental}
Christopher~A Lipinski, Franco Lombardo, Brian~W Dominy, and Patrick~J Feeney.
\newblock Experimental and computational approaches to estimate solubility and permeability in drug discovery and development settings.
\newblock {\em Advanced drug delivery reviews}, 64:4--17, 2012.

\bibitem{kabsch1976solution}
Wolfgang Kabsch.
\newblock A solution for the best rotation to relate two sets of vectors.
\newblock {\em Acta Crystallographica Section A: Crystal Physics, Diffraction, Theoretical and General Crystallography}, 32(5):922--923, 1976.

\bibitem{du2011hergcentral}
Fang Du, Haibo Yu, Beiyan Zou, Joseph Babcock, Shunyou Long, and Min Li.
\newblock hergcentral: a large database to store, retrieve, and analyze compound-human ether-a-go-go related gene channel interactions to facilitate cardiotoxicity assessment in drug development.
\newblock {\em Assay and drug development technologies}, 9(6):580--588, 2011.

\bibitem{harris2023benchmarking}
Charles Harris, Kieran Didi, Arian~R. Jamasb, Chaitanya~K. Joshi, Simon~V. Mathis, Pietro Lio, and Tom Blundell.
\newblock Benchmarking generated poses: How rational is structure-based drug design with generative models?, 2023.

\end{thebibliography}
\end{document}